\documentclass[usegraphicx,usenatbib]{mn2e}
\newcommand\ion[2]{#1$\;${\scshape{#2}}}%
\newcommand{\nH}{\ensuremath{n_\mathrm{H}}}
\newcommand{\Hn}{\ensuremath{\mathrm{HI}}}
\newcommand{\Hp}{\ensuremath{\mathrm{HII}}}
\newcommand{\Hen}{\ensuremath{\mathrm{HeI}}}
\newcommand{\Hep}{\ensuremath{\mathrm{HeII}}}
\newcommand{\Hepp}{\ensuremath{\mathrm{HeIII}}}
\newcommand{\logNH}{\ensuremath{\log(N_\mathrm{H}/\mathrm{cm}^{-2})}}

\title[QSO I-Front Thickness and the SED]{The Thickness of High-Redshift
  Quasar Ionization Fronts as a Constraint on the Ionizing Spectral
  Energy Distribution}

\author[R. H. Kramer and Z. Haiman]{R. H. Kramer$^{1}$\thanks{E-mail:
roban@astro.columbia.edu (RHK); zoltan@astro.columbia.edu (ZH)} and Z. Haiman $^{1}$\footnotemark[1]\\
$^{1}$Department of Astronomy, Columbia University, 550 West 120th
Street, New York, NY 10027}

\begin{document}

\date{Submitted to MNRAS}

\pubyear{2007}

\maketitle

\label{firstpage}

\begin{abstract}
  High--redshift quasars ($z\ga 6$) drive ionization fronts into the
  intergalactic medium (IGM), with the thickness of the front
  generally increasing with the hardness of ionizing spectrum. If the
  thickness of the front can be measured, it can provide a novel
  constraint on the ionizing spectral energy distribution (SED),
  uniquely available for sources prior to the end of reionization.
  Here we follow the propagation of an I--front into a uniform IGM,
  and compute its thickness for a range of possible quasar spectra and
  ages, and IGM neutral hydrogen densities and clumping factors.  We
  also explore the effects of uniform and non-uniform ionizing
  backgrounds. We find that even for hard spectra, the fronts are
  initially thin, with a thickness much smaller than the mean free
  path of ionizing photons.  The front gradually thickens as it
  approaches equilibrium in $10^8-10^9$ years, and the thickness of
  its outer part can eventually significantly exceed simple estimates
  based on the mean free path.  With a sufficiently high intrinsic
  hydrogen column density obscuring the source ($\logNH \ga 19.2$)
  or a sufficiently hard power--law spectrum combined with some
  obscuration (e.g. $d\log F_\nu/d\log\nu\ga -1.2$ at $\logNH \ga
  18.0$), the outer thickness of the front exceeds $\sim 1$ physical
  Mpc and may be measurable from the three--dimensional morphology of
  its redshifted 21cm signal.  We find that the highly ionized inner
  part of the front, which may be probed by Lyman line absorption
  spectra, remains sharp for bright quasars unless a large obscuring
  column ($\logNH \ga 19.2$) removes most of their ionizing photons
  up to $\approx 40 \: \mathrm{eV}$.  For obscured sources with
  $\logNH \ga 19.8$, embedded in a significantly neutral IGM, the
  black Lyman $\alpha$ trough (where the neutral fraction is $\sim
  10^{-3}$) underestimates the size of the \ion{H}{ii} region by a
  factor of $\ga 4$.
\end{abstract}


\section{Introduction}

The spectra of $z>6$ quasars
\citep[e.g.][]{Fan_etal2002,White_etal2003}, showing complete
Gunn--Peterson (GP) absorption, have been widely analyzed as a
possible probe of the ionization state of the ambient IGM.  The sizes
of inferred \ion{H}{ii} regions surrounding these sources depend on the
neutral fraction \citep{CH2000,MR2000} and suggest a value above a few
percent \citep{Fan_etal2002,WL2004,MH2004}, although this inference is
subject to some bias and significant scatter, depending on how the
intrinsic \ion{H}{ii} region size is deduced from the observed spectrum
\citep{Maselli_etal2007,BH2007b,BH2007a}.  Direct evidence for a GP
damping wing in the QSO spectra \citep{MH2004,MH2007} also suggests
that the neutral fraction exceeds several percent at least along two
lines of sight. The evolution of Lyman $\alpha$ emitting galaxies may
also suggest a significant global neutral fraction
(\citealt{Ota_etal2007}, although the presently observed evolution may
still be attributed to the evolution in the halo abundance and in the
mean IGM density; \citealt{DWH2007}).

If the IGM is significantly neutral, then the shape of the I--front
will be well--defined, i.e. the ionizing flux, over a region
corresponding to the mean--free--path of ionizing photons, will be
dominated by the quasar itself, rather than by the background. In this
case, the shape of the I--front will contain information on the
spectrum of the ionizing source.  In particular, the thickness of the
front is generally expected to increase for harder source spectra,
whose typical ionizing photon has a longer mean free path (hereafter
m.f.p.). \citet{MH2004} noted that the onset of the Lyman $\alpha$ and
$\beta$ GP troughs in the spectra of $z>6$ quasars occur at redshifts
close to one another, corresponding to a physical separation of
$R_\beta-R_\alpha\la 1$ Mpc along the line of sight, possibly
placing an upper limit on the hardness of the spectrum (with mean
photon energy $\langle E \rangle < 230$ eV, for which the m.f.p. in a
neutral IGM at $z\approx 6$ is $\sim 1$ Mpc).  Although the ratio of
oscillator strengths of the Lyman $\alpha$ and $\beta$ lines is a
factor of 5.2, once density inhomogeneities and foreground Lyman
$\alpha$ absorption is taken into account, the Lyman $\beta$ line
effectively probes a factor of $2-3$ lower neutral fraction than Lyman
$\alpha$ \citep{Fan_etal2002, SC2002}.  The comparison of $R_\beta$
and $R_\alpha$ gives the path--length over which the neutral fraction
rises by a factor of $2-3$; this could naively be taken as a proxy for
the thickness of the I--front.

This simple interpretation, however, is complicated by possible
absorption by the GP damping wing from the IGM \citep{MH2004, MH2007},
and also by the scatter in the $R_\beta/R_\alpha$ ratio expected to
arise between different lines of sight due to density fluctuations
\citep{BH2007b}.  Nevertheless, if a sufficient number of distant
quasars are detected in the future, the ratio $R_\beta/R_\alpha$, and
other features in their absorption spectra, could provide a diagnostic
for a finite I--front thickness and therefore the hardness of the sources,
at least statistically.  Note that this test is not available at lower
redshift, where the IGM is highly ionized, the m.f.p. of ionizing
photons is already $\ga 100$ Mpc, and photoionization equilibrium is
established on a time--scale of $(\Gamma/\nH
\lambda_\mathrm{mfp})^{-1} \sim 10^4$ years, much less than the
expected quasar lifetime.

Other methods have been proposed to probe the finite thickness of
quasar I--fronts.  \citet{ZS2005} explored measuring the thickness of
I--fronts through their three--dimensional redshifted 21cm signatures,
with a low-frequency radio telescope array such as LOFAR \citep[for a
comprehensive review of current and planned high-redshift 21-cm
projects, see][]{FOB_review_2006}. The soft spectrum of stellar
radiation would produce a much sharper ionization front than the very
hard spectrum predicted to emerge from miniquasars; when the sources
themselves are not detected, the 21cm maps could still discriminate
between stellar-- and quasar--driven reionization \citep{Oh2001}.
Finally, \citet{Cantalupo_etal2007} recently suggested that
collisionally excited Ly$\alpha$ emission may be detectable from the
shell of material at the I--front, over an extended solid angle behind
bright quasars; the spectral shape and thickness of this emission is
another possible measure of the thickness of the I--front.

At present, little is known empirically about the ionizing spectrum of
quasars at $z>6$. At lower redshift, quasars are found to have
relatively soft power-law spectra in the extreme ultraviolet, with an
average spectral index of around $s=1.8$ \citep{Telfer_etal2002}.
However, many quasars show a strong soft X--ray excess at a few 100eV,
which can not be explained by thermal emission from a standard
accretion disk.  The excess X-rays could originate from Compton
scattering in a hot corona surrounding the accretion disk around the
AGN \citep{Porquet_etal2004}.  Quasars at high redshift are also
expected to be preferentially obscured due to the increased gas
density, and may effectively radiate only at energies in soft X--rays
bands and above \citep{SOS2004}.  \citet{Shemmer_etal2006} found that
a composite X-ray spectrum of 21 $z>4$ quasars had an effective
spectral index in the X--ray of $s=0.95$, and an intrinsic absorption
column upper limit of $N_\mathrm{H} \la 6 \times 10^{22} \:
\mathrm{cm}^{-2}$, consistent with the mean results from
\citet{Just_etal2007} at lower redshift.

Motivated by the suggestions above that the I--front thickness could
be measurable, and by the lack of knowledge about the SEDs of
high--$z$ quasars, in this paper we follow the time--dependent
propagation of an I--front into the IGM, and study its thickness for a
range of possible spectra, quasar ages, clumping factors in the IGM,
and various ionizing backgrounds.  Our goal is to give a rough
quantitative assessment of the conditions under which a hard spectrum
may be diagnosed in future observations.  Our modeling here extends
the investigation of \citet{ZS2005}, by exploring a range of quasar
spectral parameters, by improving the treatment of helium and of
secondary ionizations by energetic photoelectrons, and, most
importantly, by following the time--dependent evolution of the
non--equilibrium ionization structure. In an earlier study,
\citet{SIR2004} used numerical simulations to study the evolution and
properties of I--fronts propagating into the high--$z$ IGM, for three
different source spectra appropriate for massive population II and III
stars and quasars. Our paper extends this work by including
recombination radiation from helium, and by focusing on brighter
sources and covering a larger range of spectra (including obscured
spectra much harder than those examined by \citeauthor{SIR2004}), for
which the thickness may be directly measurable.

After we completed this paper, we became aware of the work of
\citet{TZ2007} who have recently made calculations similar to some of
those presented here. We will compare our results further in
\S~\ref{sConclusions}.

The rest of this paper is organized as follows. 
In \S~\ref{sModel}, we describe our modeling of the I--fronts, including
treatments of the non--equilibrium evolution, and of helium.
In \S~\ref{sEvolution}, we describe the salient features -- evolution,
shape and size -- of the ionization front in our fiducial model for an
obscured source with a hard spectrum.
In \S~\ref{sThickness}, we present our main results, discussing
measurements of the thickness of the I--Front.
In \S~\ref{sCaveats}, we reiterate the inherent limitations of our
modeling approach, before finally, in \S~\ref{sConclusions} we
summarize our results and conclusions.

\section{Modeling and Methods}\label{sModel}

In this section, we describe our model for computing the propagation
of the I--front into a uniform medium.  The modeling is rather
standard for the most part, but we include a description of our own
numerical implementation here for completeness, and in order to
describe a few details about helium ionization that have typically not
been included in previous works studying cosmological quasar
\ion{H}{ii} regions.

\subsection{Ionization and Recombination}\label{ssIonization}

\citet{OsterbrockAndFerland} give an excellent description of
one-dimensional spherical ionization structure calculations. For the
most part we follow their methods, but we additionally include time evolution
(non-equilibrium ionization states), secondary ionization by
photoelectrons, and a full treatment of doubly-ionized helium.

We use the equations of one-dimensional radiative transfer to find the
optical depth $\tau(\nu)$ as a function of distance from the quasar
$r$ and frequency $\nu$. The optical depth contribution from each
species is calculated at its ionization threshold:
\begin{equation}\label{optical_depth_intergral}
\tau_{i,0}(r) = \int_0^r  \sigma_{0,i}\:n_i(r')\:dr'
\end{equation}
where $n_i(r)$ is the number density of species $i$ and $\sigma_{0,i}$
is the photoionization cross section at the threshold frequency
$\nu_{0,i}$.\footnote{Threshold ionization cross sections were taken
  from \citet{OsterbrockAndFerland} for \ion{H}{i} and \ion{He}{ii} and
  from \citet{HS1998} for \ion{He}{i}. At other frequencies, we used
  the fitting formulae from \citet{YSD1998} for \ion{He}{i} and from
  \citet{OsterbrockAndFerland} for H-like ions.} The optical depth
contribution at any frequency is then $\tau_i(\nu) =
\tau_{i,0}\:\sigma_{i}(\nu) / \sigma_{0,i}$ where $\sigma_i(\nu)$ is
the cross section of species $i$ at frequency $\nu$. The total optical
depth is the sum of contributions from all three species
\begin{equation}
\tau(\nu) = \tau_\mathrm{HI}(\nu) + \tau_\mathrm{HeI}(\nu) + \tau_\mathrm{HeII}(\nu)
\end{equation}. 

We ignore the effects of light travel time in calculating the
ionization structure. As \citet{CH2000} have pointed out, this ends up
predicting the structure that is \textit{observed} along the line of
sight. Because of the finite speed of light, the evolution of the gas
at a distance $r$ from the quasar is retarded by the interval
$r/c$. But an observer farther along the line of sight will also
receive photons from that point earlier by an amount $r/c$. So in
order to correctly predict the profile as it would be observed along
the line of sight (for instance, in an absorption spectrum), we should
add the two effects, which simply cancel each other out
\citep{White_etal2003}. If we wanted to predict the transverse
structure as it would be observed, we could simply add in the
appropriate delays after the evolution is calculated
\citep{WLB2005,Yu2005,SIA2006}.

For simplicity, we also ignore Hubble expansion. Even at $z=6$, a
quasar lifetime of $10^8$ years is much shorter than the Hubble
time. We do note, however, that the changing density of the IGM could
make a discernible difference to the shape of the ionization profile,
especially for the largest or thickest \ion{H}{ii} regions, which extend to
several percent of the Hubble radius.

The photoionization rate of species $i$ (per atom) is 
\begin{equation}
\Gamma_i = \int_{\nu_{0,i}}^\infty d\nu \: \frac{L_\nu}{h \nu} \:
\sigma_i(\nu) \: \frac{e^{-\tau(\nu,r)}}{4 \pi r^2}
\end{equation}
where $L_\nu$ is the specific luminosity, and $r$ is the distance from
the source. In our numerical ionization code, we tabulate $\Gamma_i$
as a function of $\tau_{\Hn,0}$, $\tau_{\Hen,0}$, and $\tau_{\Hep,0}$
before calculating the evolution of the ionization structure. Added to
that rate for hydrogen is a position-independent background
photoionization rate, parametrized by the resulting
equilibrium ionized fraction $X_{\Hp, BG}$:  
\begin{equation}
\Gamma_{BG} = C \alpha_B \nH\:\frac{X_{\Hp, BG}^2}{1-X_{\Hp, BG}}
\end{equation}
where $\alpha_B$ is the so-called ``case B'' recombination rate (the
recombination coefficient to all excited levels of hydrogen, evaluated
at $T_\mathrm{gas} = 10^4 \:\mathrm{K}$ unless otherwise specified)
and $C$ is the clumping factor $C \equiv \langle \nH^2 \rangle/\langle
\nH \rangle^2$.  This background arises from preexisting galaxies, and
is expected to consist of a patchwork of \ion{H}{ii} bubbles
surrounded by a relatively neutral IGM.  Note that even if the
ionization outside the quasar's \ion{H}{ii} region is patchy (with a
swiss--cheese topology), as it is expected to be, in the interior of
the \ion{H}{ii} region, where the low neutral fraction results in a
long m.f.p., it will be much more uniform (and will also have a higher
amplitude).  There will, however, be a radial profile to the
background flux, due to the clustering of galaxies around the quasar
\citep{WL2007, LMZ2007, AA2007}.  We will first ignore these effects,
assuming that the I--front is located sufficiently far away that this
bias is small, and that we are averaging the I-front over many small
galaxy--bubbles. However, in \S~\ref{sssBG}, we will return to this
issue, where we will also explore various models for a non--uniform
ionizing background.

In addition to the quasar's radiation and the UV background, several
other processes occur in the gas that affect the ionization balance
and couple the ionization states of hydrogen and helium.

Photoelectrons produced during ionization carry residual energy that
can cause further ionizations \citep{ShullSteenberg1985}. In our
calculations we include the secondary ionization of hydrogen by
photoelectrons from both hydrogen and helium, and treat it as an
on-the-spot process (the latter assumption is justified similarly to
the argument for case-B recombination, since the collisional
ionization cross--section is typically larger than the
photo--ionization cross--section; \citealt{ShullSteenberg1985}).  We
use the fitting formula provided by \citet{DHL2004} in the high-energy
limit to calculate the fraction of energy $\phi(X_e)$ each
photoelectron will expend in further ionizations of hydrogen. Using
the high-energy limit introduces only a small error since low-energy
photons cause few secondary ionizations in any case and the function
quickly approaches its asymptotic value as the photon energy
increases. Note that in the original calculations of $\phi(x)$,
\citet{ShullSteenberg1985} defined $x \equiv n_\Hp / \nH$ and assumed
$n_\mathrm{He} / \nH = 0.1$ (versus $0.079$ in our model), and $X_\Hep
= X_\Hp$ (where $X_\Hep \equiv n_\Hep / n_\mathrm{He}$). While neither
of those conditions holds exactly in our models, they are reasonable
approximations.  To account approximately for the extra electrons
introduced by doubly-ionized helium (which was not included
originally) we use $x = X_\Hp + (n_\mathrm{He}/n_\mathrm{H}) X_\Hepp$
(where $X_\Hepp \equiv n_\Hepp / n_\mathrm{He}$). It is a small effect
in any case, since \Hepp{} is formed only where hydrogen is already
highly ionized, and $\phi$ vanishes as $x$ approaches unity.

We neglect secondary ionizations of helium which happen at less than
$20\%$ of the rate of hydrogen. The mean number of secondary
ionizations of hydrogen per photoelectron produced in the ionization
of species $i$ is
\begin{equation}
n_2 = \phi(X_e)\:\frac{\langle E_i \rangle - h \nu_{0,i}}{h \nu_{0,\Hn}}
\end{equation}
where $\langle E_i \rangle$ is the mean energy of photons locally
ionizing species $i$:
\begin{equation}\label{eMeanE}
\langle E_i \rangle = \frac{1}{\Gamma_i} \int_{\nu_{0,i}}^\infty d\nu \: (h \nu) \:
\sigma_i(\nu) \: \frac{L_\nu}{h \nu} \: \frac{e^{-\tau(\nu,r)}}{4 \pi
  r^2}
\end{equation}
Similarly to $\Gamma_i$, we tabulate $\langle E_i \rangle$ as a
function of $\tau_{\Hn,0}$, $\tau_{\Hen,0}$, and $\tau_{\Hep,0}$,
before integrating the ionization structure. 

Photons produced during recombination of hydrogen and helium are also
included in our code in the on-the-spot approximation. High-frequency
recombination radiation is capable of ionizing neutral hydrogen and
helium, as well as singly-ionized helium, so the fraction of
recombination photons ultimately ionizing each species must be
calculated. In general the fraction of photons of frequency $\nu$ that
ionize species $i$ is
\begin{equation}
y_{i}(\nu) = \frac{\sigma_i(\nu)\:n_i}{\sigma_{\Hn}(\nu)\:n_{\Hn} +
  \sigma_{\Hen}(\nu)\:n_{\Hen} + \sigma_{\Hep}(\nu)\:n_{\Hep}}
\end{equation}
The overall fraction for a given recombination process is an average
of $y_i(\nu)$ over the spectrum of the emitted radiation. In our
calculations, we make the simplifying assumptions that all \ion{H}{i},
\ion{He}{i} and \ion{He}{ii} Lyman continua, and \ion{He}{ii} Balmer
continuum photons are produced with the minimum energy. The first and
last of these occur below the threshold for all species but \Hn{}. The
\ion{He}{ii} Lyman $\alpha$ line is treated similarly, assuming all
photons have the mean energy of that line.

Certain higher \ion{He}{i} lines produces in the recombination cascade
are energetic enough to ionize hydrogen, with an average of 0.96
H-ionizing photons ultimately produced for each recombination to any
excited level of \Hen{} \citep[in the low-density limit which applies
here,][]{OsterbrockAndFerland}. Another source of ionizing photons is
the transition from the first excited level to the ground state of
\Hep{}: instead of producing a \ion{He}{ii} Lyman $\alpha$ photon, it
can instead occur via a two-photon process. We calculate the fraction
of recombinations to all excited states of \Hep{} that end in a
two-photon emission (as a function of temperature) by interpolating
the tabulated values from \citet{HS1964}. An average of 1.425 photons
capable of ionizing hydrogen and 0.737 capable of ionizing neutral
helium are produced for each two-photon event \citep{FP1980}, the
latter of which are assumed to have the threshold ionization energy for
helium.

We use recombination rate coefficients from \citet{HG1997}, except for
the coefficient for recombinations directly to the $n=2$ level of
\Hep, which is taken from \citet{FF1997}. All calculations presented
in this paper assume an isothermal gas of constant mean density and
clumping factor.

\subsection{The Ionization Structure}

The ionization structure surrounding the quasar is calculated by
integrating the net ionization rates over time. Before the first
iteration, an array of ($X_{\Hp}$, $X_{\Hep}$, $X_{\Hepp}$) values,
corresponding to a sequence of $r$ values, is initialized to
($X_{\Hp,BG}$, 0, 0). The ionization and recombination rates are
calculated as described above and used to find $dX_{\Hp}/dt$,
$dX_{\Hep}/dt$, and $dX_{\Hepp}/dt$. The integration time step is
calculated using
\begin{equation}
 \Delta t = \frac{F_{dt}}{\max( \Gamma_{\Hn} X_{\Hn},~ \Gamma_{\Hen}
 X_{\Hen},~ \Gamma_{\Hep} X_{\Hep} )}
\end{equation}
where $F_{dt} = 0.004$ is just a numerical parameter optimized by
trial and error. Decreasing $F_{dt}$ by a factor of two (improving the
time resolution) has no important effect on the results. The time step
is only updated every 10 iterations, and to help damp numerical
oscillations we actually use $\Delta t_\mathrm{new} = 0.75~\Delta
t_\mathrm{old} + 0.25~\Delta t$ as the time step.

In order to increase the speed and stability of the integration, the
ionized fractions for the inner part of the array (at low $r$) are
frozen and no longer recalculated once a stable equilibrium is
achieved. Every tenth iteration we calculate the maximum radius at
which the ionization states are in equilibrium, defined as $r_\mathrm{eq} =
\max(r)\:\mathrm{for~which}$
\begin{eqnarray}
X_{\Hn} \left( \frac{dX_{\Hp}}{dt} \right ) ^{-1} > 5 \times 10^7 \: \Delta t\\
X_{\Hep} \left( \frac{dX_{\Hepp}}{dt} \right ) ^{-1} > 5 \times 10^7
\: \Delta t\\
\tau_{\Hn,0} < 1000
\end{eqnarray}
The last condition on the optical depth is introduced to exclude
regions far ahead of the front, where time scales grow long because
the radiation field is so weak. For runs with a high uniform ionizing
background we decreased the optical depth limit to $\tau_{\Hn,0} <
100$ because of the reduced opacity. Once $r_\mathrm{eq}$ is found,
the next iteration is begun, and the ionized fractions at
$r>r_\mathrm{eq}$ are updated (and constrained to the interval
$[0,1]$). Using our fiducial parameters (described below and in \S~\ref{sEvolution}), it took 76000 iterations to evolve the structure
through $5 \times 10^8$ years.

\subsection{Numerical Parameters}

As mentioned above, ionization rates and mean photon energies are
tabulated as a function of the threshold optical depths of each
species. The quantities are evaluated on a grid of logarithmically
spaced values of $\tau_{\Hn,0}$, $\tau_{\Hen,0}$, and
$\tau_{\Hep,0}$. The dimensions of the ionization rate grids are 332
by 301 by 301 values, while the mean photon energy grids are 416 by 61
by 61 values. Each optical depth range extends from $10^{-8}$ up to
the maximum optical depth for which $\Gamma_i/\max(\Gamma_i) >
10^{-6}$, which depends on the source spectrum.

The simulations presented here use a grid of 1300 equally spaced
radius values, extending from $5 \times 10^4 \:\mathrm{pc}$ (all
distances in this paper are in proper, not comoving, units unless
stated otherwise) to $1.5 \: R_\mathrm{Strom}$, where
$R_\mathrm{Strom}$ is the radius of the classic equilibrium
Str\"omgren sphere,
\begin{equation}
R_\mathrm{Strom} \equiv \left( \frac{Q}{\frac{4}{3} \pi \nH{}^2
  \alpha_B C}\right)^\frac{1}{3}.
\end{equation}
Note that this expression ignores helium and secondary
ionizations. With $Q = 2 \times 10^{57} \: s^{-1}$ and $C=1$,
$R_\mathrm{Strom} = 22.2 \: \mathrm{Mpc}$ at $z=6$.

We tested for convergence by doubling the resolution, which had no
effect on the results. For our canonical parameters the step size
works out to $\Delta r = 2.56 \times 10^{-2} \:\mathrm{Mpc}$ with an
outer radius of $33.4 \:\mathrm{Mpc}$.

We have tested our code several ways to ensure reasonable numerical
performance. As mentioned above, we tested for convergence by doubling
either the time or spatial resolution, neither or which had any
important effect on the results. Using a soft spectrum (so that the
front is sharp) we find very good agreement with analytical estimates
for ionized region size and front velocity. We also find agreement
with the Cosmological Radiative Transfer Codes Comparison Project
\citep[``test 1'' in][]{CRTCC} calculations, and the examples
presented in \citet{OsterbrockAndFerland}.

\subsection{Physical Parameters}\label{ssPhyicalParameters}

The temperature of the gas is important, because it controls the
recombination rates. All calculations presented in this paper assume
an isothermal gas. We justify this assumption by noting that most of
the photoionization heating occurs somewhat ahead of the front where
the ionized fractions are small and recombination rates are therefore
low, whereas the temperature inside the front is fairly uniform
\citep[as can be seen in][]{BH2007a, CRTCC}. We experimented with
introducing artificially varying temperature profiles in the outer
regions of the front. We made the temperature a function of optical
depth so that the temperature variations would be tied to the location
of the front. We tried several profiles, from smooth functions
mimicking the temperature profiles in the appendix of \citet{BH2007a},
to a step function increasing the temperature by a factor of 5 beyond
$\tau_{\mathrm{HI},0} \ge 0.001$.  We found that none of these had a
significant effect on the profile, as long as the variation occurred
outside of the region that had reached equilibrium. The ionization
rate totally dominates the recombination rate in the moving part of
the front, reaching parity only as the ionized fraction approaches
equilibrium. We conclude that since the heating of the gas occurs
before it reaches ionization equilibrium, and since the exact
recombination rate is irrelevant outside of the equilibrium region,
changes in the recombination rate due to temperature variations will
not be an important factor in determining the shapes of the
fronts.\footnote{Changes in temperature due to radiative transfer can
  also have dynamical effects on the gas around the I-front; but in
  the limit where the recombination rate is unimportant, our
  conclusions on the ionized fraction should also be insensitive to
  such dynamical effects.}

We assume the number density of hydrogen atoms surrounding our sources
is the mean density of the IGM at $z=6$, $\nH = 2.2 \times 10^{-7} \:
\mathrm{cm}^{-3} \: (1+z)^3 = 7.6 \times 10^{-5}
\:\mathrm{cm}^{-3}$. The helium mass fraction is $0.24$, yielding
$n_\mathrm{He} = 6.0 \times 10^{-6} \: \mathrm{cm}^{-3}$. The electron
temperature is $T_\mathrm{gas} = 10^4 \:\mathrm{K}$. For our fiducial
case we use a clumping factor of $C=1$.

\begin{figure}
\includegraphics[height=3in,angle=270]{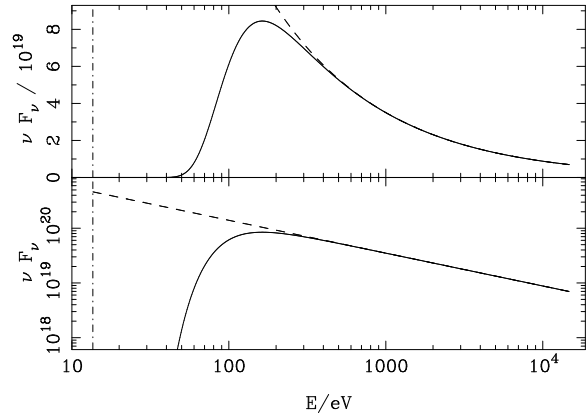}
\caption{An absorbed power law spectrum (arbitrary units) with $s =
  1.6$ and $N_\mathrm{H} = 10^{19.2} \: \mathrm{cm}^{-2}$ (solid
  curves) plotted in linear (upper panel) and logarithmic (lower
  panel) flux units. The dashed lines show the unabsorbed spectrum. The
  vertical dot-dashes lines indicate the hydrogen ionization threshold
  at $13.6 \: \mathrm{eV}$.}\label{fSpectrum}
\end{figure}

The spectra we have used are absorbed power laws, as show in Figure
\ref{fSpectrum}, with specific luminosity
\begin{equation}
L_\nu \propto \nu^{-s} \exp \left( - N_\mathrm{H} \left[ \sigma_{\Hn}(\nu)
+ \frac{n_\mathrm{He}}{n_\mathrm{H}} \sigma_{\Hen}(\nu) \right] \right),
\end{equation}
where $s$ is the spectral index and $N_\mathrm{H}$ and $N_\mathrm{He}$
are the column densities of hydrogen and helium intrinsic to the
source. The spectrum is normalized to produce ionizing photons at rate
$Q$ before the absorption, so the ionizing photon luminosity after
absorption $Q_\mathrm{abs}$ can be much lower. Note that high energy
photons can cause multiple secondary ionizations, so $Q_\mathrm{abs}$
is not exactly the total ionization rate the spectrum produces. It is
difficult to take secondary ionizations into account \textit{a
  priori}, since (as we discussed earlier) the number of secondary
ionizations is a strong function of the local free electron density,
and therefore depends on where the photon is absorbed. Other factors
can have an even larger effect on the ionizing efficiency of the
spectrum, such as the fact that the hardest photons can escape the
region entirely.

As mentioned in the Introduction, the ionizing spectra of quasars at
$z>6$ is uncertain.  For our fiducial model for an obscured source, we
have chosen $Q=2\times 10^{57} \: \mathrm{s}^{-1}$ (corresponding
roughly to the expected value for the $z\sim 6$ quasars in the SDSS;
e.g. \citealt{MH2004}), $s = 1.6$, and $\logNH = 19.2$ (which results
in a post-absorption photon luminosity of $Q_\mathrm{abs} = 8.3 \times
10^{55} \: \mathrm{s}^{-1}$).  We will consider variations in the
power--law index and obscuring column density that are consistent with
expectations based on the lower redshift data.

\section{Evolution,  Shape and Size of the Ionization Front}\label{sEvolution}

In this section, we present our results for the time--evolving
I--front around a fiducial source with hard spectrum, and discuss the
broad features that can be relevant to future measurements.
Quantitative results will be presented and discussed in the next
section.

\subsection{The Time-Evolving Ionization Front}\label{ssEvolution}

An often used proxy for the thickness of an ionization front around a
quasar is the mean free path $l_\mathrm{MFP} = \left[ ~ n_{\Hn}
  \:\sigma_{\Hn}(\langle \nu \rangle) ~ \right]^{-1}$, where $\langle \nu
\rangle$ is a characteristic frequency of the ionizing photons. The
thickness can be estimated, for example, by adopting $n_{\Hn} = 0.5 \:
\nH$, and the mean frequency of ionizing photons emitted by the quasar
for $\langle \nu \rangle$. The resulting value of $l_\mathrm{MFP}$,
however, is an accurate proxy for the thickness of the I--front only for
particular definitions of the thickness, and only under particular
conditions. For instance, the ionizing spectrum is hardened as the
photons travel outward through the gas, since low-energy photons are
more readily absorbed. As a result, the outer edge of the front will
be spread over a longer path than the inner edge.

Another complication is that the ionization fronts around short--lived
quasars are expected to be propagating outward (rather than
corresponding to the static edge of an equilibrium Str\"omgen sphere),
which has a strong effect on the shape of the front. It is widely
noted (e.g., \citealt{SG1987}) that the ionization state of the gas
surrounding a typical high-redshift quasar will not have time to reach
equilibrium with the radiation field during a quasar lifetime of
$10^7$ to $10^8$ years,\footnote{A combination of overdensity and high
  clumping factors in the typical quasar environment could, however,
  reduce the recombination time to a value perhaps as short as $\sim 5
  \times 10^6$ years at $z=6.4$ \citep{YuLu2005}.}
since the time scale to reach equilibrium is roughly the recombination
time $t_\mathrm{rec} = (C\,\alpha_B\,\nH)^{-1} = 1.62 \times 10^9 \, C^{-1}\,
\left[(1+z)/7\right]^{-3} \: \mathrm{years}$.  This timescale can also
be thought of as the time required for the source to emit one photon
for each hydrogen atom within the Str\"omgren sphere. Throughout the
process of establishing equilibrium, the shape of the ionization front
is changing, therefore we must take into account non-equilibrium
effects in order to correctly predict the ionization front thickness.

\begin{figure}
\includegraphics[height=3in,angle=270]{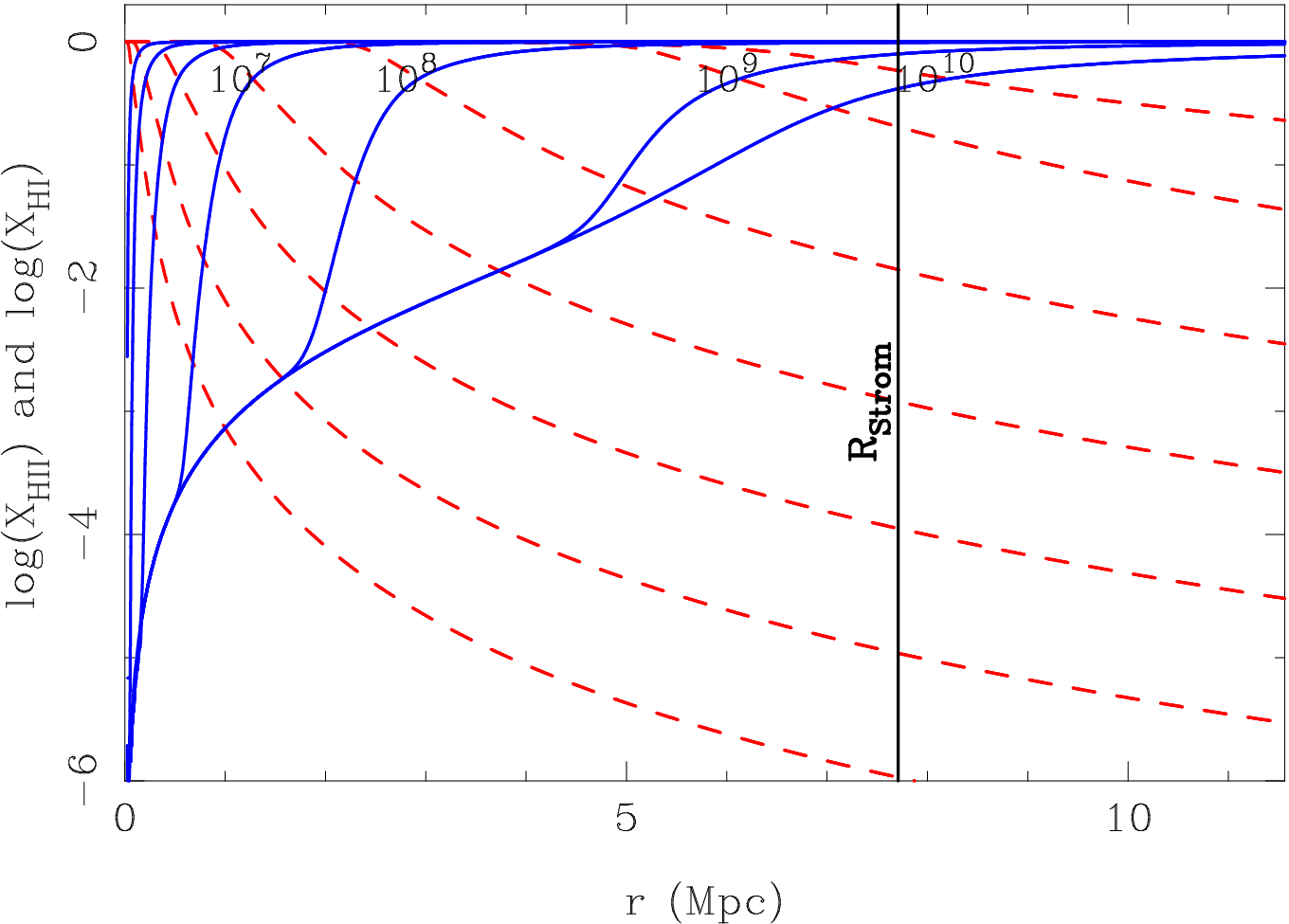}\\
\includegraphics[height=3in,angle=270]{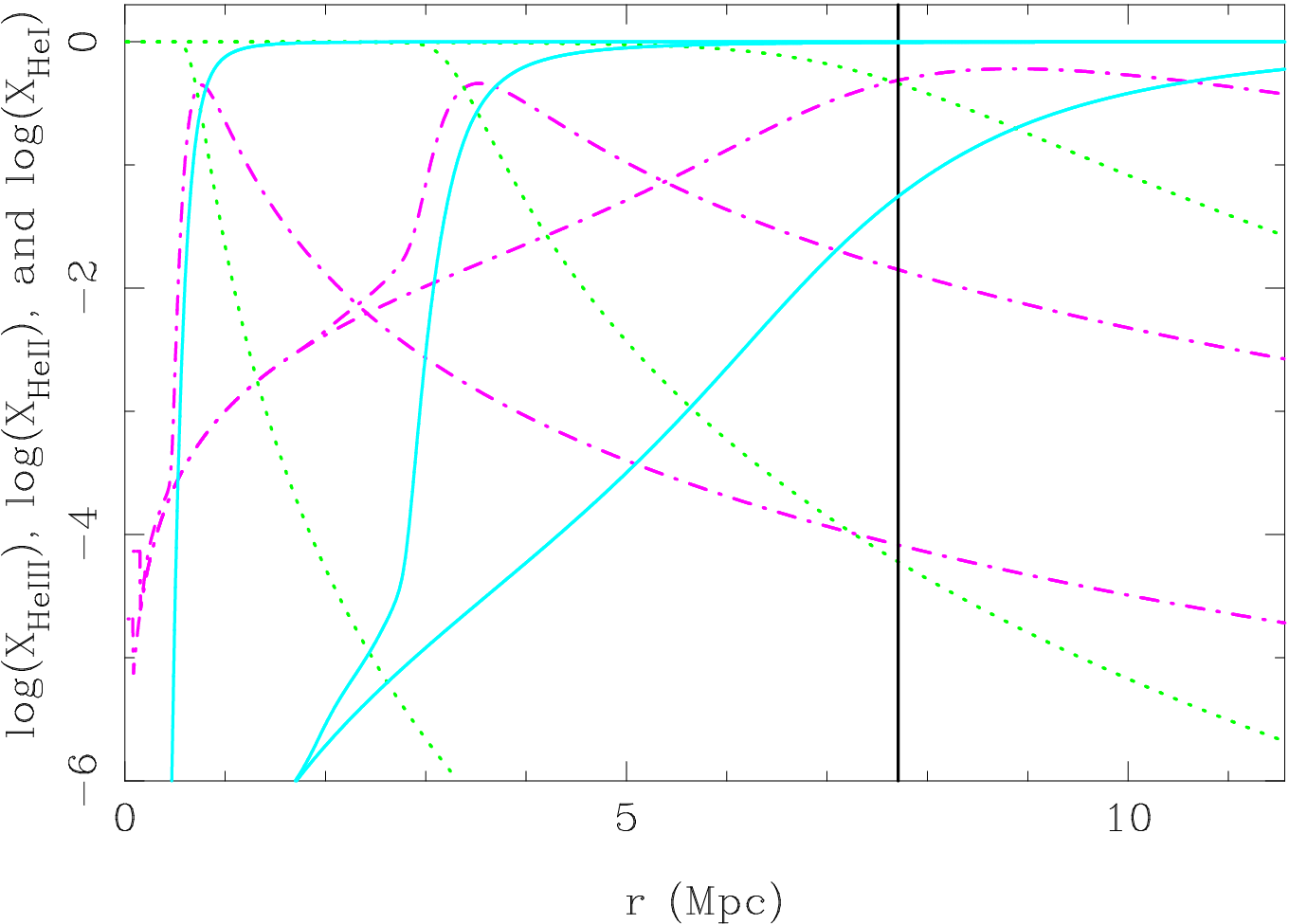}
\caption{Ionized (dashed red) and neutral (solid blue) hydrogen
  fractions (a); and neutral (solid cyan), singly-ionized (dot-dashed
  magenta) and doubly-ionized (dotted yellow) helium fractions (b)
  versus radius (proper coordinates). Curves in the upper are for
  quasar ages of $4\leq \log(t/\mathrm{years})\leq 10$ in 1 dex
  increments. In the lower panel, only $\log(t/\mathrm{years}) = 6$,
  $8$, and $10$ are shown for clarity. By $10^{10}$ years the entire
  ionization structure is in equilibrium. Model parameters are
  $Q=2\times 10^{57} \: \mathrm{s}^{-1}$; $s = 1.6$; $\logNH = 19.2$.
  The mean photon energy for this spectrum is $\approx 240$ eV, and
  the corresponding mean free path is $\sim 3$ (proper) Mpc.}
  \label{fFrontEvolution}
\end{figure}

Figure \ref{fFrontEvolution} shows the evolution of the ionization
structure around our fiducial absorbed power--law source.  The
hydrogen front starts out quite thin, but becomes thicker and thicker
as it propagates outward. When the source is turned on, the gas
closest to it is very quickly ionized up to its equilibrium level,
while gas a little further away is ionized much more slowly, due both
to the intervening absorption and geometric dilution of the
radiation. This means that the inner edge of the front is evolving
faster than the outer edge, which effectively tilts the $X_{\Hn}$
curve, making it steeper. This creates a very thin transition region
at first, which thickens as the front propagates outward and slows
down. Depending on how the thickness of the front is measured, it can
start out thinner than the mean free path, and end up much thicker.

\begin{figure}
\includegraphics[height=3in,angle=270]{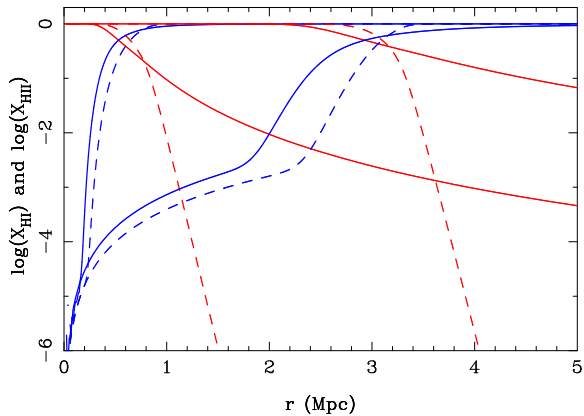}
\caption{Ionization structure for an absorbed power law spectrum
  (solid, same parameters as Figure \ref{fFrontEvolution}) and for a
  monochromatic spectrum (dashed). Ionized (red, decreases with $r$)
  and neutral (blue, increases with $r$) hydrogen fractions are shown
  at $t=10^6\:\mathrm{years}$ (left set of curves) and
  $t=10^8\:\mathrm{years}$ (right set of curves). The monochromatic
  spectrum is normalized to produce ionizing photons at the same rate
  $Q_\mathrm{abs} = 8.3 \times 10^{55} \: \mathrm{s}^{-1}$ and has a
  photon energy of $89 \: \mathrm{eV}$, which is the local mean
  ionizing photon energy (weighted by cross section) $\langle
  E_\mathrm{H} \rangle$ of the absorbed power law spectrum. Note that
  a convenient way to estimate the thickness of the front is to
  compare where the ionized and neutral fraction curves cross $\log(X)
  \approx -1$ at a given time.}
  \label{fFrontMono}
\end{figure}

\begin{figure}
\includegraphics[height=3in,angle=270]{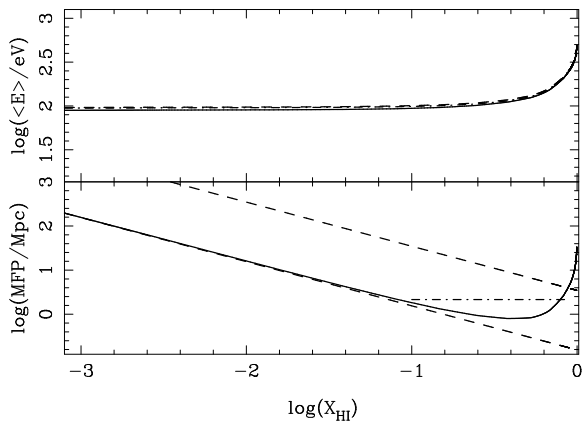}
\caption{The upper panel shows the mean ionizing photon energy
  $\langle E_\mathrm{H} \rangle$ for hydrogen (solid curve), neutral
  helium (dashed curve), and singly-ionized helium (dot-dashed curve).
  The lower panel shows the local mean free path versus neutral
  hydrogen fraction at $10^8$ years (solid curve; using the same model
  as in Figure \ref{fFrontEvolution}). The lower dashed line shows
  $l_\mathrm{MFP}$ calculated using the $\tau=0$ value for the local
  mean ionizing photon energy (weighted by cross section) $\langle
  E_\mathrm{H} \rangle = 89 \: \mathrm{eV}$, while the upper dashed
  line is calculated using the $\tau=0$ overall mean ionizing photon
  energy $\langle E \rangle = 239 \: \mathrm{eV}$. The horizontal
  dot-dashed line indicates the thickness of the front as measured
  from $X_{\Hn} = 0.1$ to $0.9$. }
  \label{fMeanE}
\end{figure}

These effects can be seen even with a monochromatic spectrum, but, as
illustrated in Figure \ref{fFrontMono}, a spectrum with a broad
continuum of ionizing radiation can enhance the effect. Early on
(e.g. at $10^6$ years), photons with lower energy (and therefore high
ionization cross section) will dominate the ionization at the front,
while higher-energy photons escape to larger distances and pre-ionize
the medium (as seen in the long tail of curves representing the
ionized fraction). This makes the front relatively thin because of the
short mean free path of these photons. As the front moves outward (e.g
at $10^8$ years), the lowest energy photons will continue to be
preferentially absorbed closer to the source (and will dominate the
ionization balance in the inner regions), leaving the higher-energy
photons to dominate in the outer parts of front. This hardened
spectrum creates a thicker front because of the longer mean free
path. 

Based on the above features, it will be useful for the discussion
below to divide a snapshot of the ionization structure at a time $t <
t_\mathrm{rec}$ into three distinct regions: Closest to the quasar, in
the {\it equilibrium zone}, the ionization rate has already reached
equilibrium with the recombination rate. Just outside the equilibrium
region is a {\em rapid--ionization zone}, where relatively neutral gas
is suddenly being exposed to intense ionizing radiation. Further out
there is a {\em pre--ionization zone}, where the relatively rare
high-energy photons penetrate and begin to ionize the gas, but
recombination rates are low because of the low ionized fraction.

Figure \ref{fMeanE} shows a snapshot of the local mean ionizing photon
energy (equation \ref{eMeanE}), and the corresponding mean free path
(calculated using the local neutral fraction), versus neutral hydrogen
fraction, for the profile at $10^8$ years shown in
Figure~\ref{fFrontEvolution}. The spectrum is relatively unaffected by
hardening up to $X_{\Hn} \approx 0.1$, but the mean energy increases
rapidly after the neutral fraction reaches $0.3$. Similarly, the mean
free path at first decreases in inverse proportion to the increasing
neutral fraction, but reaches a minimum at $X_{\Hn} \approx 0.3$ and
begins to grow as the spectrum hardens. This explains why the outer
tail of ionization is so thick compared to the inner face of the
front. If we measure the thickness of the front from $X_{\Hn} = 0.1$
to $0.9$, we find (as shown in the figure) that it is barely
consistent with the range of $l_\mathrm{MFP}$ over the same interval:
both approaches to estimating the $l_\mathrm{MFP}$ based on an average
photon energy of the source become inaccurate because of the spectral
hardening. The intersection of the three curves at
$\log\left(X_{\Hn}\right) = -0.1$ is merely a coincidence. Indeed,
recall that the ionization structure is not in equilibrium, so the
width is constantly changing. The estimate of the m.f.p. based on the
mean photon energy of $239\:\mathrm{eV}$ and a neutral medium yields
$\sim 3$ Mpc, whereas the path corresponding to the range $0.1 <
X_{\Hn} < 0.9$ in the equilibrium case is much longer, $\sim 15$
Mpc.

\begin{figure}
\includegraphics[height=3in,angle=270]{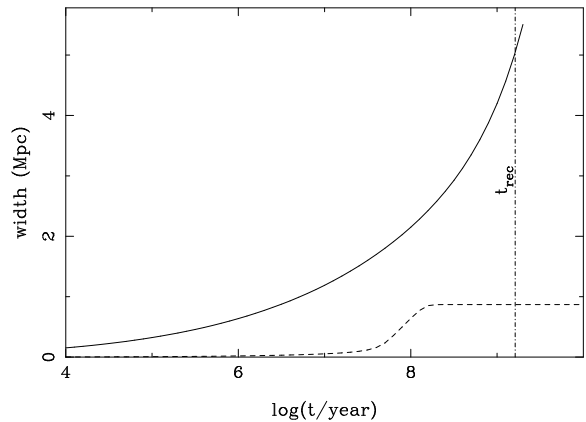}
\caption{Width measurements versus time for the ionization front shown
  in Figure \ref{fFrontEvolution}. The solid curve shows the thickness of
  the front as measured from $X_{\Hn} = 0.1$ to $0.9$, while the
  dashed curve shows the thickness between $X_{\Hn} = 10^{-3}$ and
  $10^{-2.5}$.}
  \label{fWidthlogt}
\end{figure}

Figure \ref{fWidthlogt} shows two different measurements of the thickness
of the front versus the logarithm of the time. The definitions of
these thicknesses will be justified in \S~\ref{sThickness}, but for now we
note that the outer part of the front grows thicker all the way up to
about $t_\mathrm{rec}$, while the inner part of the front comes into
equilibrium much sooner.


\begin{figure}
\includegraphics[height=3in,angle=270]{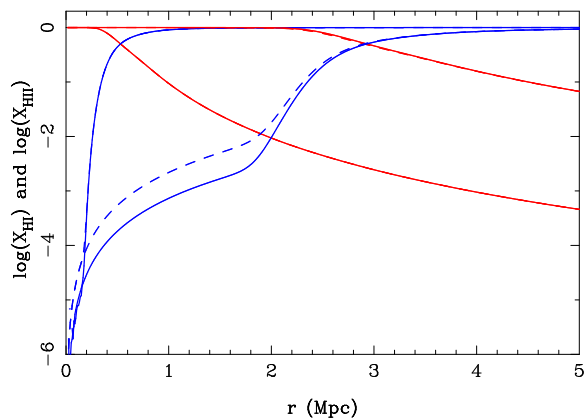}
\caption{Ionization structure with (solid) and without (dashed)
  ionization of hydrogen by helium recombination photons. Ionized
  (red, decreases with $r$) and neutral (blue, increases with $r$)
  hydrogen fractions are shown at $t=10^6\:\mathrm{years}$ (left set
  of curves) and $t=10^8\:\mathrm{years}$ (right set of curves). Note
  that this recombination radiation accounts for almost all of the
  influence of helium on the hydrogen ionization structure.}
  \label{fFrontnoRR}
\end{figure}

\begin{figure}
\includegraphics[height=3in,angle=270]{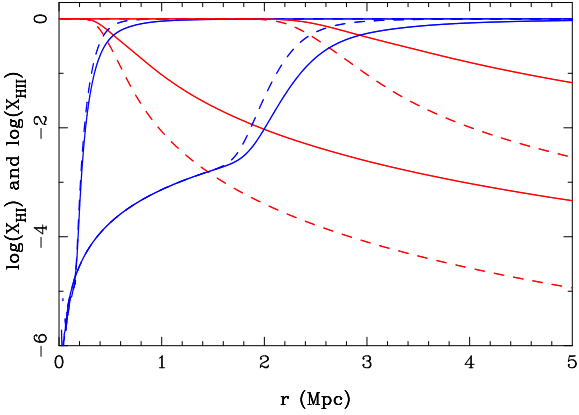}
\caption{Ionization structure with (solid) and without (dashed)
  secondary ionizations by high energy photoelectrons. Ionized (red,
  decreases with $r$) and neutral (blue, increases with $r$) hydrogen
  fractions are shown at $t=10^6\:\mathrm{years}$ (left set of curves)
  and $t=10^8\:\mathrm{years}$ (right set of curves).}
  \label{fFrontnoSI}
\end{figure}

The presence of helium also has important effects on the hydrogen
ionization structure, as can be seen in Figure~\ref{fFrontnoRR}. The
influence is due to the coupling of the ionization states by
recombination radiation. By reprocessing high-energy photons (which,
in the absence of helium, would escape to large distances) into
lower-energy recombination radiation (which is absorbed on the spot,
mostly by hydrogen), helium lowers the hydrogen neutral fraction in
the equilibrium zone.  The difference is a factor of $\sim 3$ in the
region where $X_{\Hn}\la 10^{-2}$ at $10^8$ years; this region is
relevant in the analysis of the proximity zone of the quasar in
Ly$\alpha/\beta$ absorption. In particular, neglecting helium can
result in predicting a much thicker front, depending on how thickness
is measured (and a factor of several too large $R_\beta/R_\alpha$
ratio; see below).

Secondary ionizations have a similarly important effect, but on the
outer tail of the ionization structure, as shown in
Figure~\ref{fFrontnoSI}. Neglecting secondary ionizations can reduce
the thicknesses we would predict in the outer part of the front by
several Mpc.

\subsection{The Effects of  Spectral Hardness}\label{ssHardness}

As evident from the above discussion, the shape and hardness of the
spectrum can have dramatic effects on the ionization
structure. Generally a harder spectrum will produce thicker fronts,
although as we shall see below, this is not always the case for
certain definitions of front thickness. The mean photon energy is not
enough to characterize the shape of the ionization structure it will
produce. For instance, even a relatively weak high-energy component
can produce an enhanced pre-ionization region, while a low-energy
component can increase the ionized fraction of the equilibrium region
but have little effect on the thickness of the rapid--ionization zone.
Enhancing the effect of a high-frequency component is the fact that
energetic photons can cause multiple secondary ionizations, especially
in a relatively neutral medium.

A harder spectrum will also generally produce a ``shallower'' ionized
region, in the sense that the ionized fraction will be lower in the
inner part of the ionized region since the ionization cross section is
lower for high-energy photons. At the quasar ages we are considering this
effect can be considered to affect mainly the equilibrium region.

Two parameters control the shape of the spectrum in our
models. Increasing the spectral index $s$ makes the spectrum softer,
while increasing the intrinsic hydrogen column density $N_\mathrm{H}$
cuts off more low-energy emission, hardening the spectrum. 

\begin{figure}
\includegraphics[height=3in,angle=270]{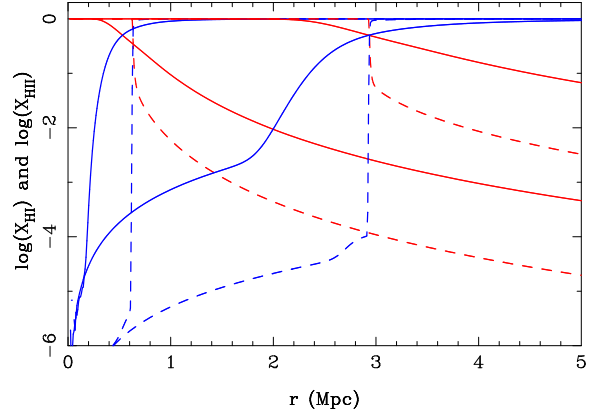}
\caption{Ionized and neutral hydrogen fractions for a source with no
  intrinsic absorption ($N_\mathrm{H} = 0$, dashed) or with $\logNH =
  19.2$ (solid). Spectra are normalized to produce the same number of
  photons $Q_\mathrm{abs} = 8.3 \times 10^{55} \:
  \mathrm{s}^{-1}$. Other spectral parameters are the same as in
  Figure \ref{fFrontEvolution}.} \label{fUnabsorbed}
\end{figure}

For reference, Figure \ref{fUnabsorbed} shows the fronts produced by a
spectrum with no intrinsic absorption ($N_\mathrm{H} = 0$). As
expected, the fronts are much sharper than those in Figure
\ref{fFrontEvolution}.

\subsection{The Size of the Ionized Region}\label{ssSize}

\begin{figure}
\includegraphics[height=3in,angle=270]{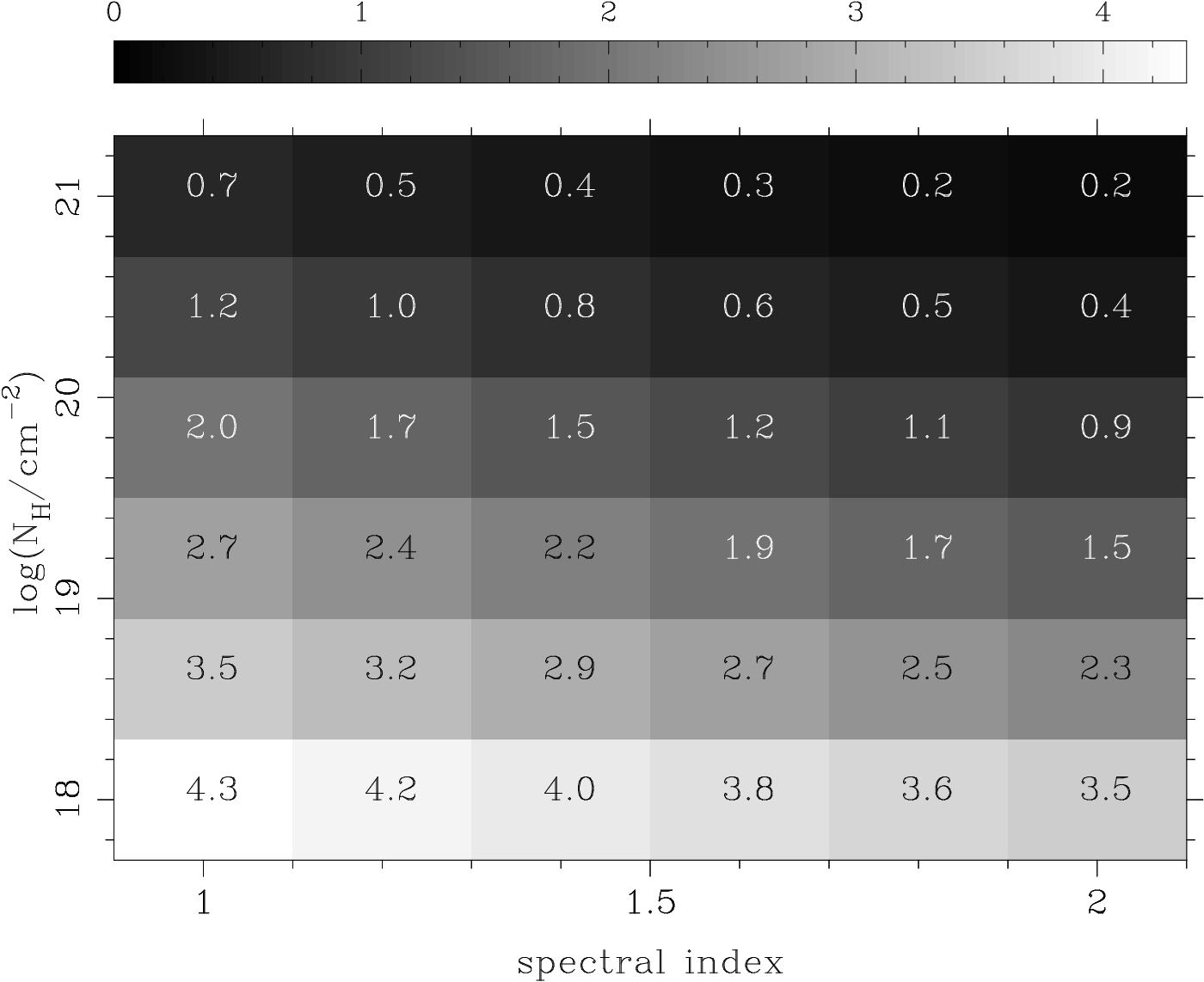}
\includegraphics[height=3in,angle=270]{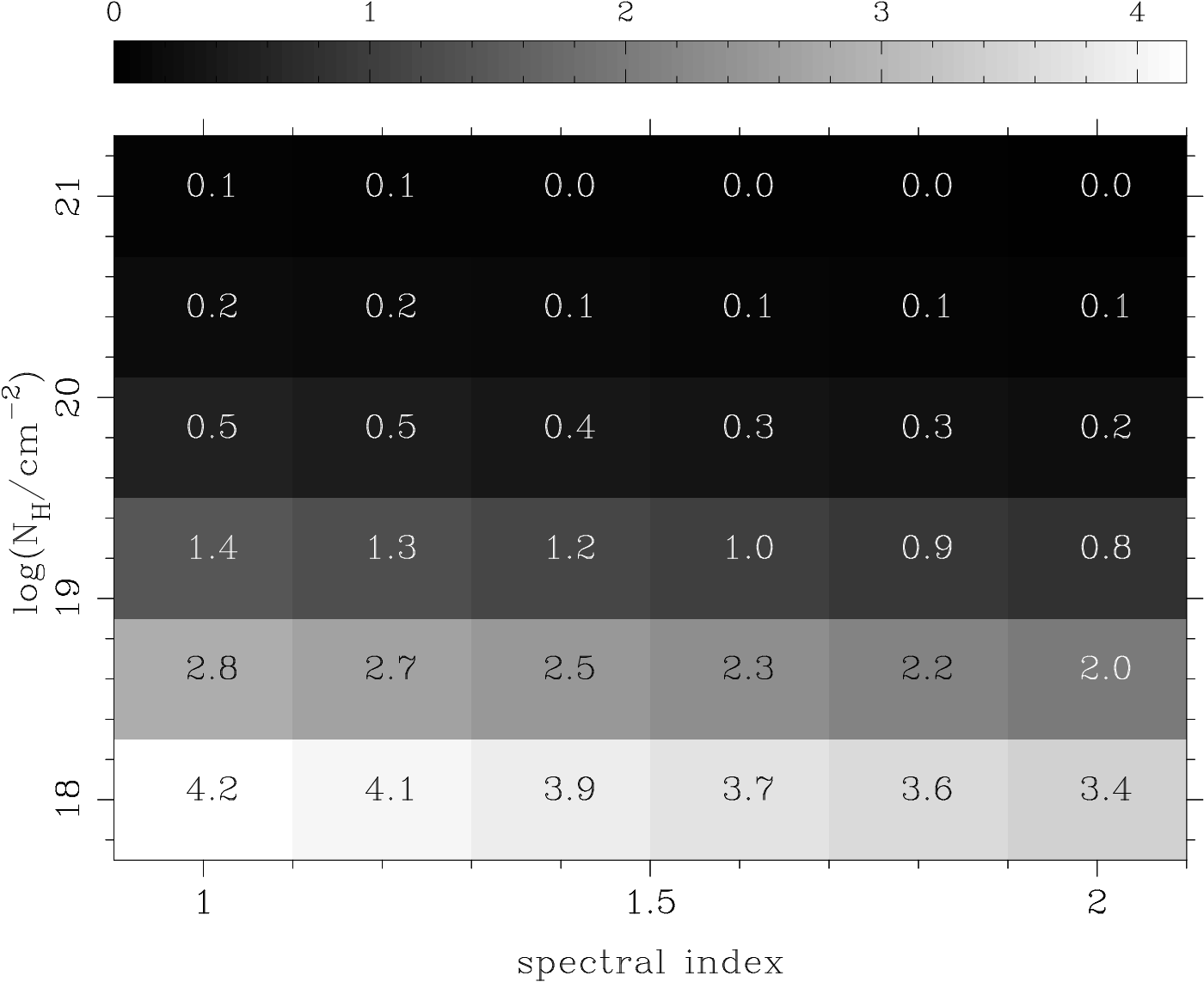}
\caption{The size of the ionized region in Mpc out to $X_{\Hn} = 0.5$
  ($r_{0.5}$, a) and $X_{\Hn} = 0.001$ ($r_{-3}$, b) versus the
  intrinsic absorption $N_\mathrm{H}$ (in $\mathrm{cm}^{-2}$) and the
  spectral index $s$ at $3.16 \times 10^7$ years and with $Q=2\times
  10^{57} \: \mathrm{s}^{-1}$. Each block represents one simulation,
  with the size indicated in numerals and by the shading. The key to
  the shading appears at the top of each graph.}\label{fSize}
\end{figure}

Before discussing the thickness of the I--fronts in detail in the next
section, it is useful to discuss the overall size of the \ion{H}{ii}
region.  As mentioned in the Introduction, the size of the \ion{H}{ii}
regions could be probed either in 21cm studies, or by Lyman line
absorption spectra of individual sources. The former technique is
sensitive to $X_{\Hn}\sim 1$, whereas the latter technique (with
typical deep Keck spectra) probes the much lower neutral fraction
$X_{\Hn}\sim 10^{-3}$.  In Figure \ref{fSize}, we therefore show the
size of the ionized region measured out to $r_{0.5}$, where $X_{\Hn} =
0.5$ (top panel), and to $r_{-3}$, where $X_{\Hn} = 10^{-3}$ (bottom
panel). The shaded plots show the \ion{H}{ii} region sizes at $3.16
\times 10^7$ years (corresponding roughly to the Salpeter time, and to
the expected ages of bright quasars; see, e.g.,
\citealt{MartiniReview}), as a function of the intrinsic absorption
$N_\mathrm{H}$ and spectral index $s$.

Remember that the softest spectrum is in the lower right corner of the
graph, while the hardest spectrum is in the upper left. The trend in
both panels is for the radius to decrease with increasing
$N_\mathrm{H}$, as $Q_\mathrm{abs}$ decreases due to the absorption,
and as more of the photons escape from the ionized region because of
their longer MFP. Decreasing the spectral index $s$ (which also
produces a harder spectrum), on the other hand, results in a larger
\ion{H}{ii} region, since the harder spectrum is less attenuated by the
intrinsic absorption, meaning $Q_\mathrm{abs}$ is larger.

Considering both \ion{H}{ii} region size measurements together leads to
an interesting conclusion. Several recent papers
\citep{MH2004,Maselli_etal2007,LMZ2007,BH2007b} have explored the
accuracy of $R_\alpha$ and $R_\beta$, defined as the radii where the
flux in the redshifted Ly$\alpha$ and Ly$\beta$ region of the spectrum
falls below some fixed threshold, as a proxy for the location of the
I--front.  They found that $R_\alpha$ and $R_\beta$ are biased
measurements, underestimating the ``true'' distance to the I--front
(which we might choose to locate at $r_{0.5}$) by typically 20--30\%.
Here we discover another effect to worry about: as the spectra get
harder (and especially as the obscuring column increases), the radius
of the highly ionized region $r_{-3}$, decreases much more quickly
than the half-ionization radius $r_{0.5}$. This is because the harder
photons are ineffective at maintaining a high ionized fraction because
of the low ionization cross section, instead tending to escape into
more neutral gas farther out. This produces a thicker front and a more
neutral \ion{H}{ii} region. For hard spectra, this significantly
increases the bias, with the underestimate reaching values as high as
a factor of $\approx 8$ for the spectra with the largest obscuring
column.
 
\section{Measuring the Thickness of the I--Front}\label{sThickness}

\subsection{Front Thickness in 21cm and in Lyman Absorption Spectra} 
\label{ssThickness}

As mentioned above, we are interested in two different regions of the
ionization front: the highly ionized inner face of the front at
$10^{-3} \la X_{\Hn} \la 10^{-2.5}$ which is approximately
the region probed by Lyman-series absorption spectra, and the
less--ionized outer tail of the front at $0.1 \la X_{\Hn}
\la 0.9$, which is the region potentially accessible to
redshifted 21 cm observations.

\citet{MH2004} proposed the idea that the difference between $R_\beta$
and $R_\alpha$, as defined in the previous section, could be useful to
probe the spectral hardness, with the two values increasingly
separated for harder spectra.  In a subsequent study, \citet{BH2007b}
simulated a large set of quasar Lyman $\alpha$ and $\beta$ absorption
spectra and studied the possibility of using the transmission window
created by the ionized region around the quasar to constrain the mean
ionization fraction of the IGM. They found that density fluctuations
induce a large scatter in their measured quantity, which is the ratio
of radii inferred from each spectrum $R_\beta/R_\alpha$.  However,
according to their simulations, $R_\beta/R_\alpha = 1.2$ would be
distinguishable from $R_\beta/R_\alpha = 1$ at the $2 \sigma$ level
with about 20 high-quality quasar spectra. As a proxy for this ratio,
we use $r_{-2.5}/r_{-3}$, the ratio of radii at which $X_{\Hn} =
10^{-2.5}$ and $X_{\Hn} = 10^{-3}$, which correspond roughly to the
points where flux would be lost in the Ly--$\beta$ and Ly--$\alpha$
absorption spectra, respectively, for a typical deep Keck spectrum of
a $z=6$ quasar.

In contrast to the Lyman line absorption spectra, the 21cm
observations will be sensitive, at $z=6-10$, to features corresponding
to neutral fractions of order unity, with regions that have
$X_{\Hn}\la 0.1$ corresponding to ``holes'' in tomographic 21cm
maps. Although the effective spatial resolution of such maps is highly
dependent on the specific interferometric instrument and its
configuration, a rough target value for next--generation facilities
may be $\sim 1$ physical Mpc (see, e.g., the review by
\citealt{FOB_review_2006}).  We adopt here $d_{0.1}$, the difference
between the radius at which $X_{\Hp} = 0.1$ and $X_{\Hn} = 0.1$, as a
proxy for the thickness of the I--front that may be measurable with
$21\:\mathrm{cm}$ instruments.

\begin{figure}
\includegraphics[height=3in,angle=270]{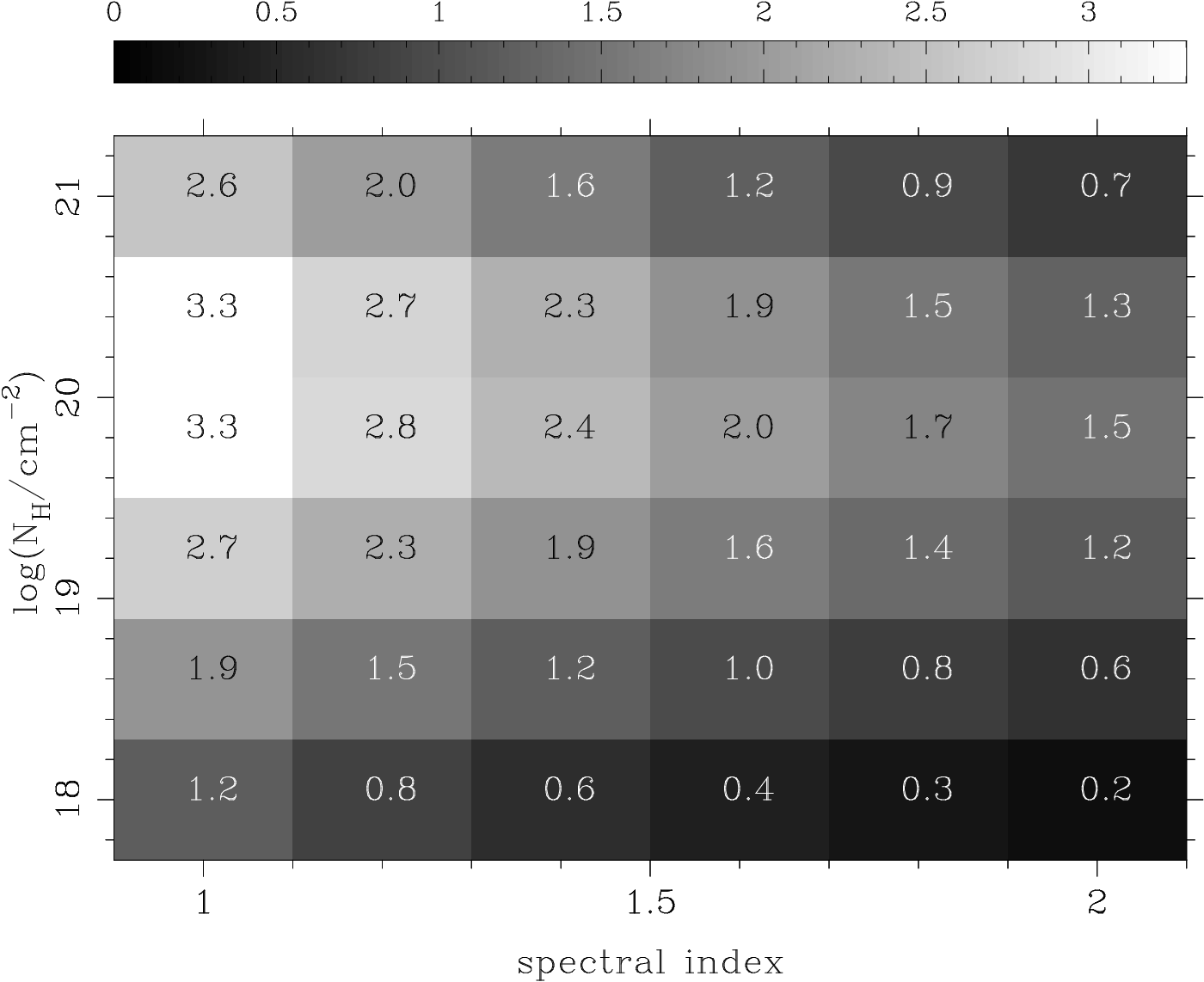}
\includegraphics[height=3in,angle=270]{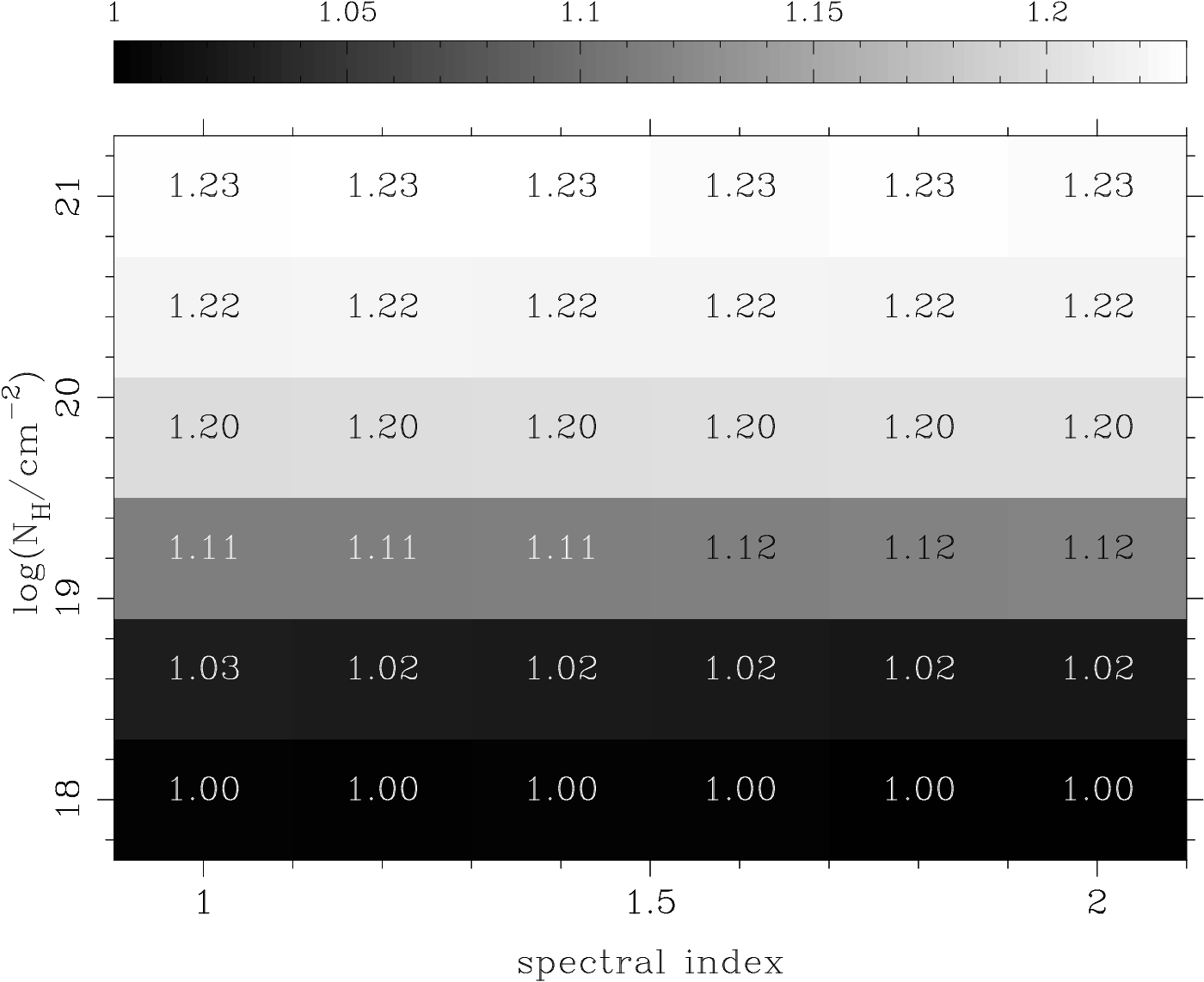}
\caption{Ionization front thickness measures at $3.16 \times 10^7$
  years as measured by $d_{0.1}$ (in Mpc, top) and $r_{-2.5}/r_{-3}$
  (bottom).}\label{f3e7}
\end{figure}

Figure \ref{f3e7} depicts our measures of the ionization front
thickness around a quasar with an age of $\sim 3.16 \times 10^7$
years, on the same grid of parameters used in Figure~\ref{fSize}.  The
first panel shows $d_{0.1}$, which exceeds our estimated detection
threshold of $\sim 1 \: \mathrm{Mpc}$ for a large region of the
parameter space. The thickness increases with increasing spectral
hardness up till $\logNH = 19.8$, after which is decreases as
$N_\mathrm{H}$ increases, due to the overall shrinking of the ionized
region discussed in \S~\ref{ssSize}, and shown in
Figure~\ref{fSize}. This will be the pattern for most of the grids we
will discuss: the $d_{0.1}$ widths will always increase with
decreasing spectral index, and will grow with increasing absorption up
to a point, but will eventually shrink for large enough
$N_\mathrm{H}$.

The $r_{-2.5}/r_{-3}$ ratio in the second panel is almost totally
insensitive to the spectral index. Making the spectral index smaller
does make the front slightly thicker, but it also moves the front
outward by a small amount, which keeps the ratio constant. The ratio
increases dramatically between $\logNH = 18.6$ and $\logNH = 19.2$,
becoming large enough to be potentially detectable (according to the
criterion defined above based on \citealt{BH2007b}), thanks to the fact
that for the harder spectra $r_{-3}$ reaches equilibrium faster, while
$r_{-2.5}$ continues to move outward a little longer, which stretches
out the ratio. The intrinsic absorption must be at least $\logNH =
19.2$ for the ratio to exceed $1.1$. The ratio exceeds $1.2$ for any
spectral index at $\logNH \ge 19.8$.

Note that the contours for the two size measures in the
$N_\mathrm{H}$-$s$ plane have different orientations. This is not
surprising, since variations of $s$ and $N_\mathrm{H}$ can affect the
profile differently at different locations in the front. This suggests
that a combination of $21 \:\mathrm{cm}$ and Lyman-series absorption
measurements could break the degeneracy inherent in the separate
observations and allow both parameters to be constrained
simultaneously.

\begin{figure}
\includegraphics[height=3in,angle=270]{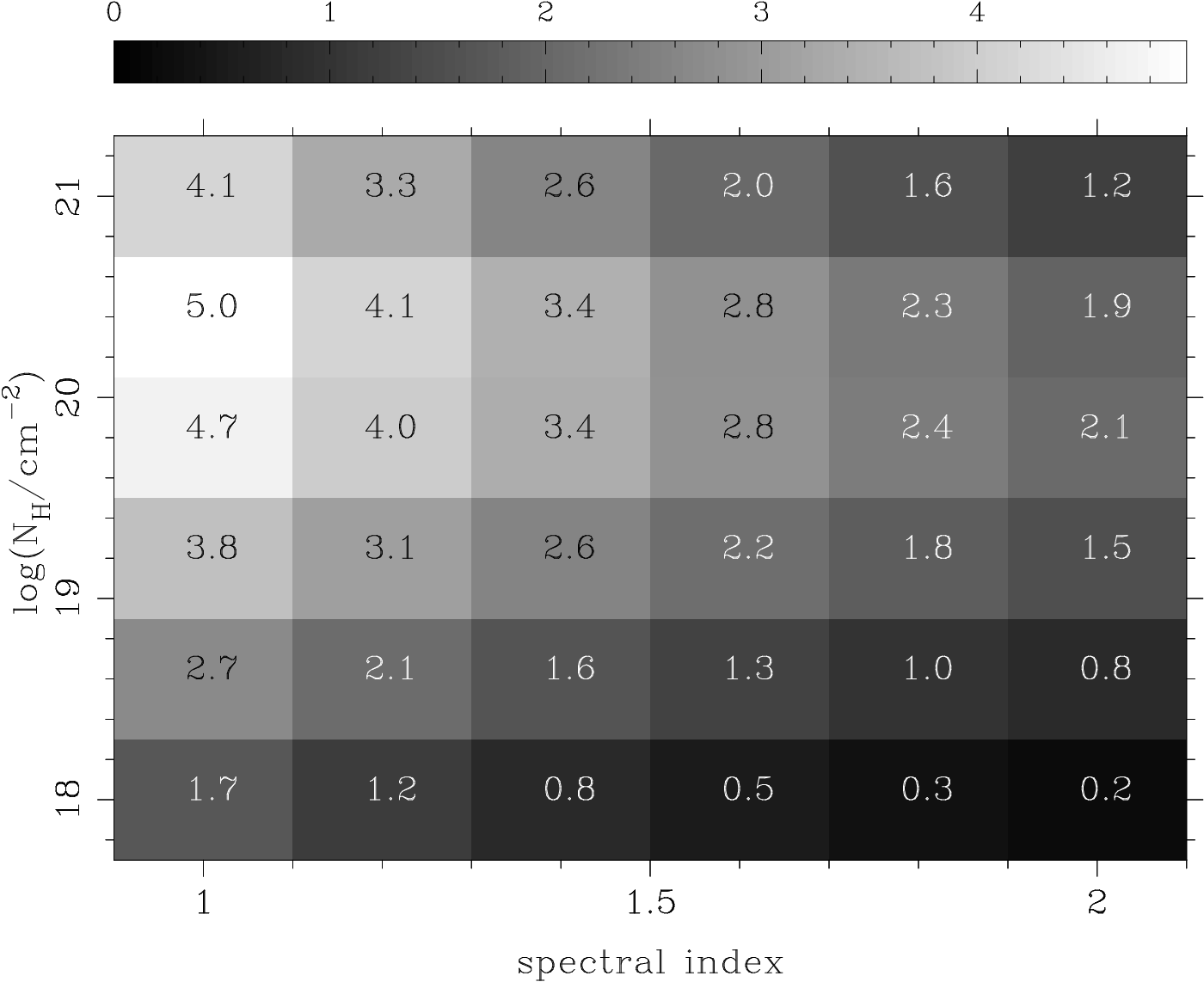}
\includegraphics[height=3in,angle=270]{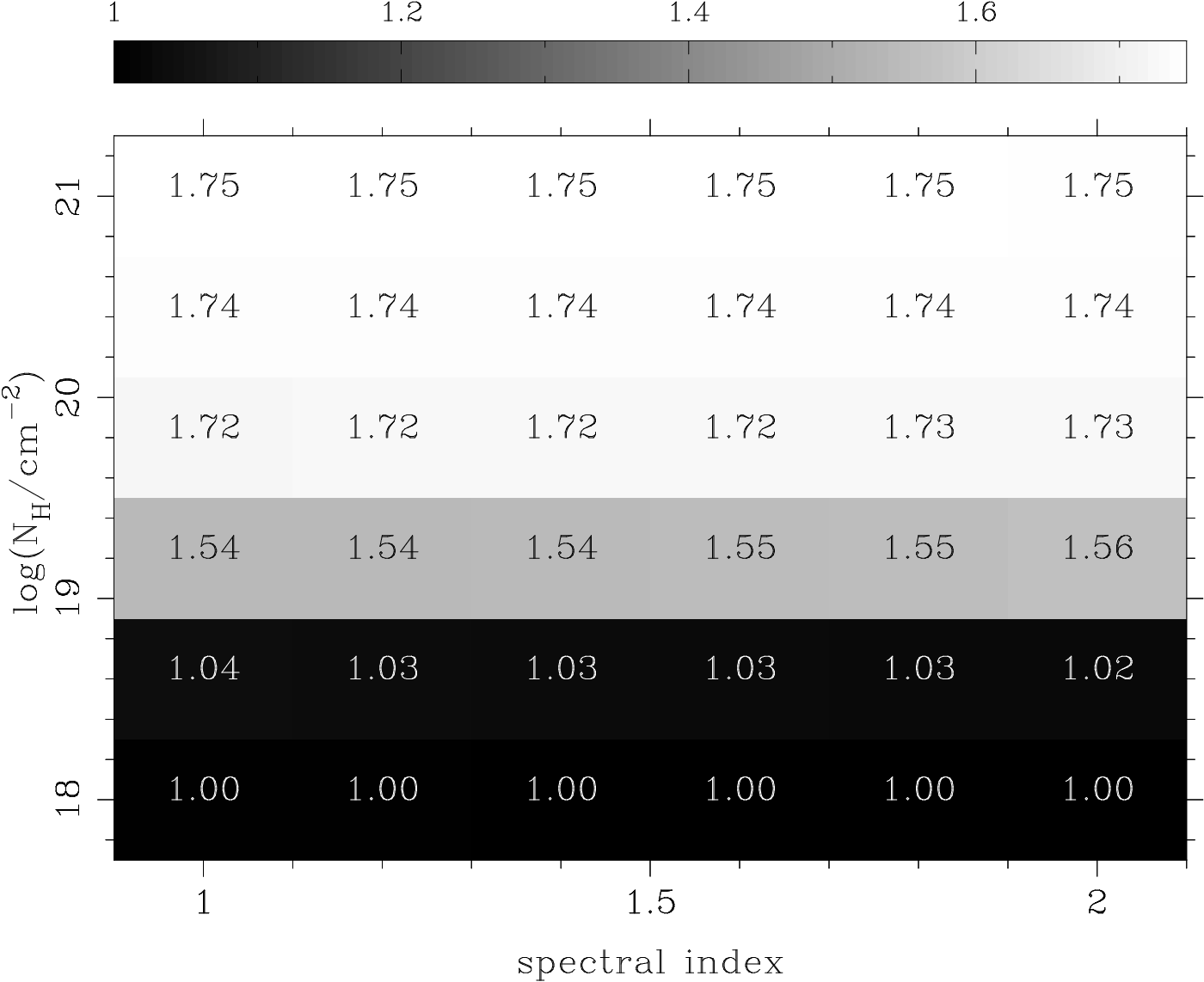}
\caption{Widths at $1 \times 10^8$ years as measured by $d_{0.1}$ (in
  Mpc, top) and $r_{-2.5}/r_{-3}$ (bottom).}\label{f1e8}
\end{figure}

\subsection{Parameter Variations}\label{ssVariations}

We next examine how our results change when we vary other model
parameters.

\subsubsection{Source Lifetime} 

In Figure \ref{f1e8}, we show the same thickness measures as in Figure
\ref{f3e7}, but for quasar three times older, at $t = 10^8$ years. The
figure reveals that the outer tail of the front ($d_{0.1}$) has
generally gotten thicker, but otherwise follows the same trends we saw
at $3.16 \times 10^7$ years. The inner face ratio $r_{-2.5}/r_{-3}$
has remained small and almost unchanged for the softest spectra with
$N_\mathrm{H} \le 10^{18.6} \: \mathrm{cm}^{-2}$. These spectra create
steep fronts and deep equilibrium regions with $X_{\Hn} < 0.001$, so
even at $10^8$ years $r_{-2.5}/r_{-3}$ is still measuring the steep
slope of the rapid ionization zone. There is a sharp increase in the
front thickness at all spectral indices between $N_\mathrm{H} =
10^{18.6}$ and $10^{19.2} \: \mathrm{cm}^{-2}$. This is because for
the harder spectra the equilibrium region reaches $X_{\Hn} > 10^{-3}$
(but is usually still below $X_{\Hn} = 10^{-2.5}$) before $10^8$ years,
meaning that $r_{-3}$ is the distance to a fixed point in the
equilibrium zone, while $r_{-2.5}$ is the distance to a point in the
still-receding ionization front. This produces the high ratios seen in
the figure.

\subsubsection{Source Luminosity} 

\begin{figure}
\includegraphics[height=3in,angle=270]{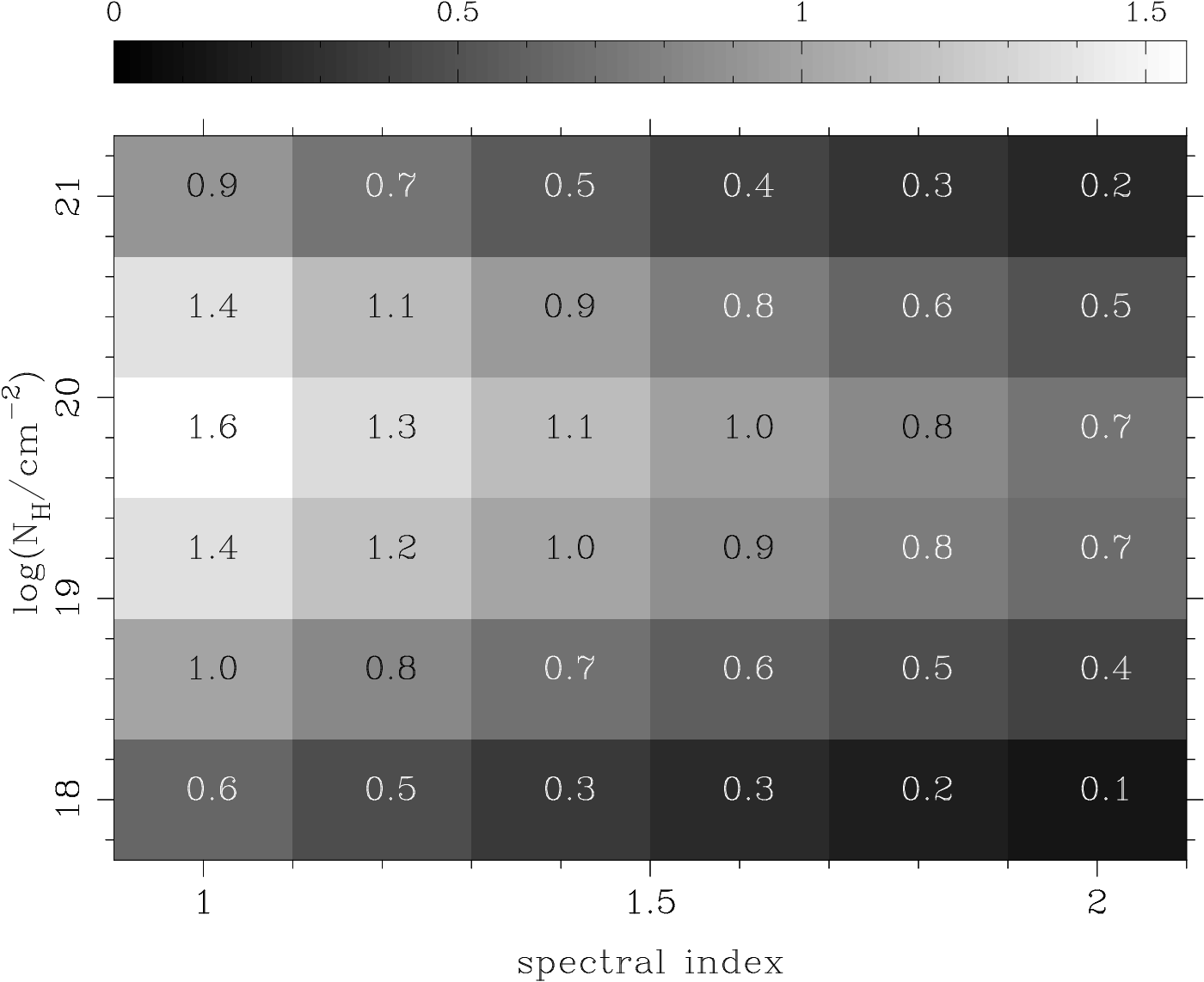}
\includegraphics[height=3in,angle=270]{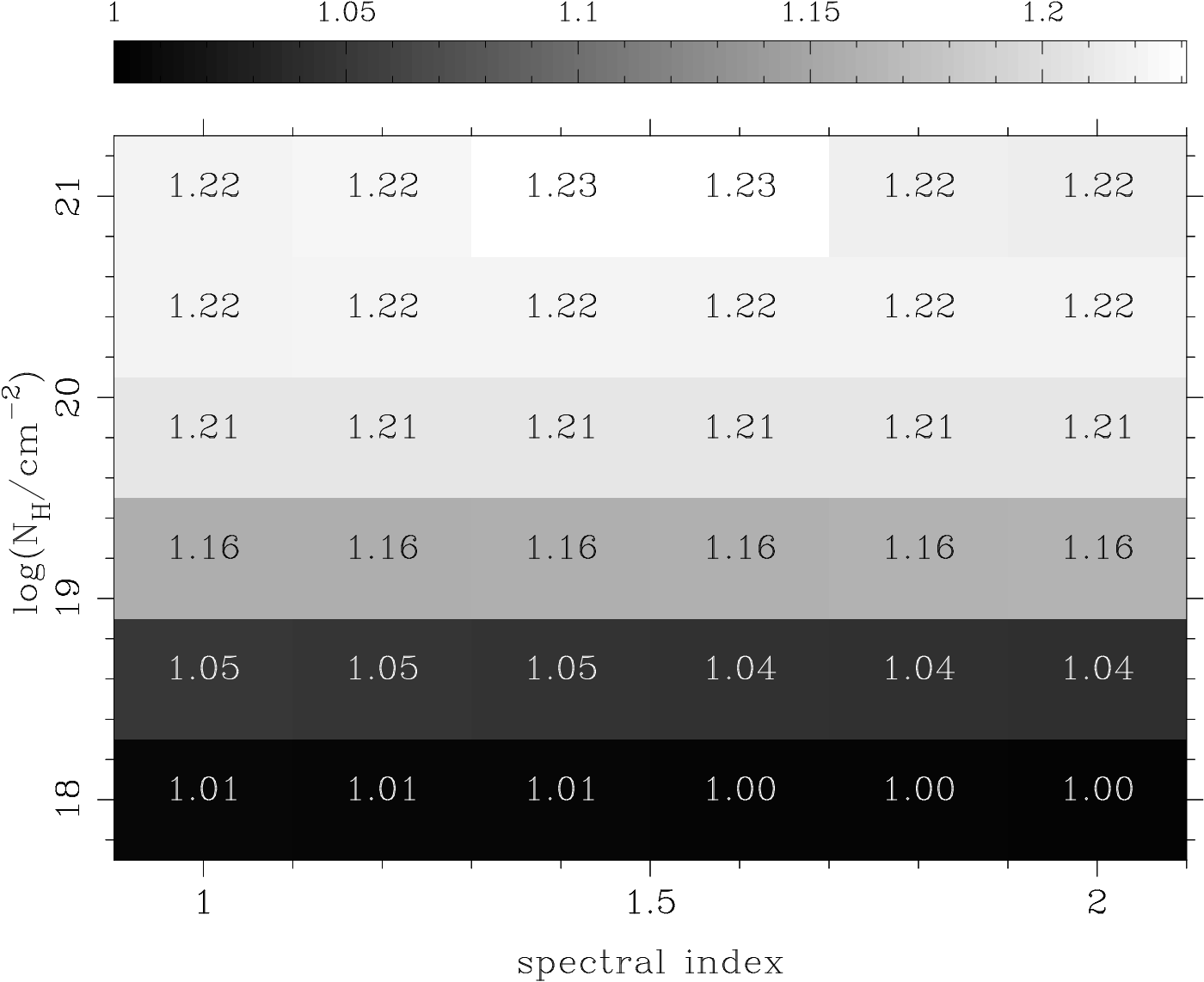}
\caption{Widths with $Q = 2 \times 10^{56} \: \mathrm{s}^{-1}$ at $3
  \times 10^7$ years as measured by $d_{0.1}$ (in Mpc, top) and
  $r_{-2.5}/r_{-3}$ (bottom).}\label{fQ56}
\end{figure}

Reducing the (pre-intrinsic-absorption) ionizing photon luminosity
from $Q = 2 \times 10^{57} \: \mathrm{s}^{-1}$ to $2 \times 10^{56} \:
\mathrm{s}^{-1}$ produces the widths shown in
Figure~\ref{fQ56}. Reducing the luminosity by a factor of $10$ reduces
the time scale by a factor of $10$ and the Str\"omgren radius by a
factor of $10^{1/3} = 2.15$. The shape of the pre-ionization and
rapid-ionization zones (since they are largely unaffected by
recombinations) are therefore almost identical to the $Q = 2 \times
10^{57} \: \mathrm{s}^{-1}$ models at $3.16 \times 10^6$ years (not
shown), when the same total number of photons has been emitted. And
indeed the $d_{0.1}$ values are very close, and consequently are lower
than the $Q = 2 \times 10^{57} \: \mathrm{s}^{-1}$ models at $3.16
\times 10^7$ years by a factor of about $1.5$ to $2$. The same
reasoning does not apply to the $r_{-2.5}/r_{-3}$ ratio, which end up
with similar values. The lower luminosity means that $r_{-3}$ is
frozen into the equilibrium zone earlier, but $r_{-2}$ also moves out
more slowly, so the ratio ends up similar to the higher luminosity
case.

\subsubsection{Clumping Factor} 

\begin{figure}
\includegraphics[height=3in,angle=270]{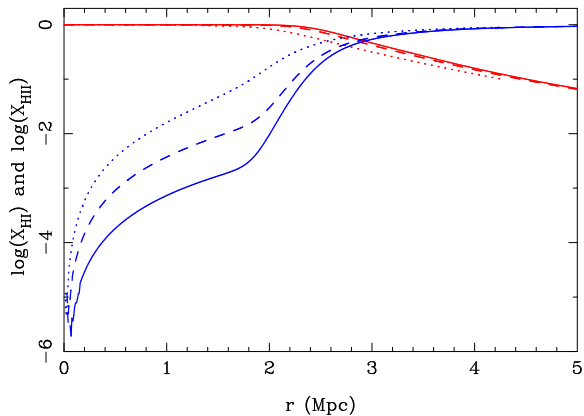}
\caption{Ionized (red) and neutral (blue) hydrogen fractions for
  $C=1,5,20$ (bottom-to-top for the neutral fraction) at $t=10^8 \;
  \mathrm{years}$. Spectral parameters are the same as in
  Figure~\ref{fFrontEvolution}.}\label{fClump}
\end{figure}

The clumping factor effectively increases the recombination
coefficient, which has the effect of raising the neutral fraction in
the equilibrium zone, but has little effect on the rapid-ionization or
pre-ionization regions, \citep[as emphasized in][]{CH2000}, and as can
be seen in Figure \ref{fClump}.  Increasing the clumping factor also
decreases the recombination time and Str\"omgren radius. With $C=5$,
$t_\mathrm{rec} = 3.2 \times 10^8 \: \mathrm{years}$, while with
$C=20$, $t_\mathrm{rec} = 8.1 \times 10^7 \: \mathrm{years}$ (both at
$z=6$), meaning that the ionization structure could approach
equilibrium within a quasar lifetime.

\begin{figure}
\includegraphics[height=3in,angle=270]{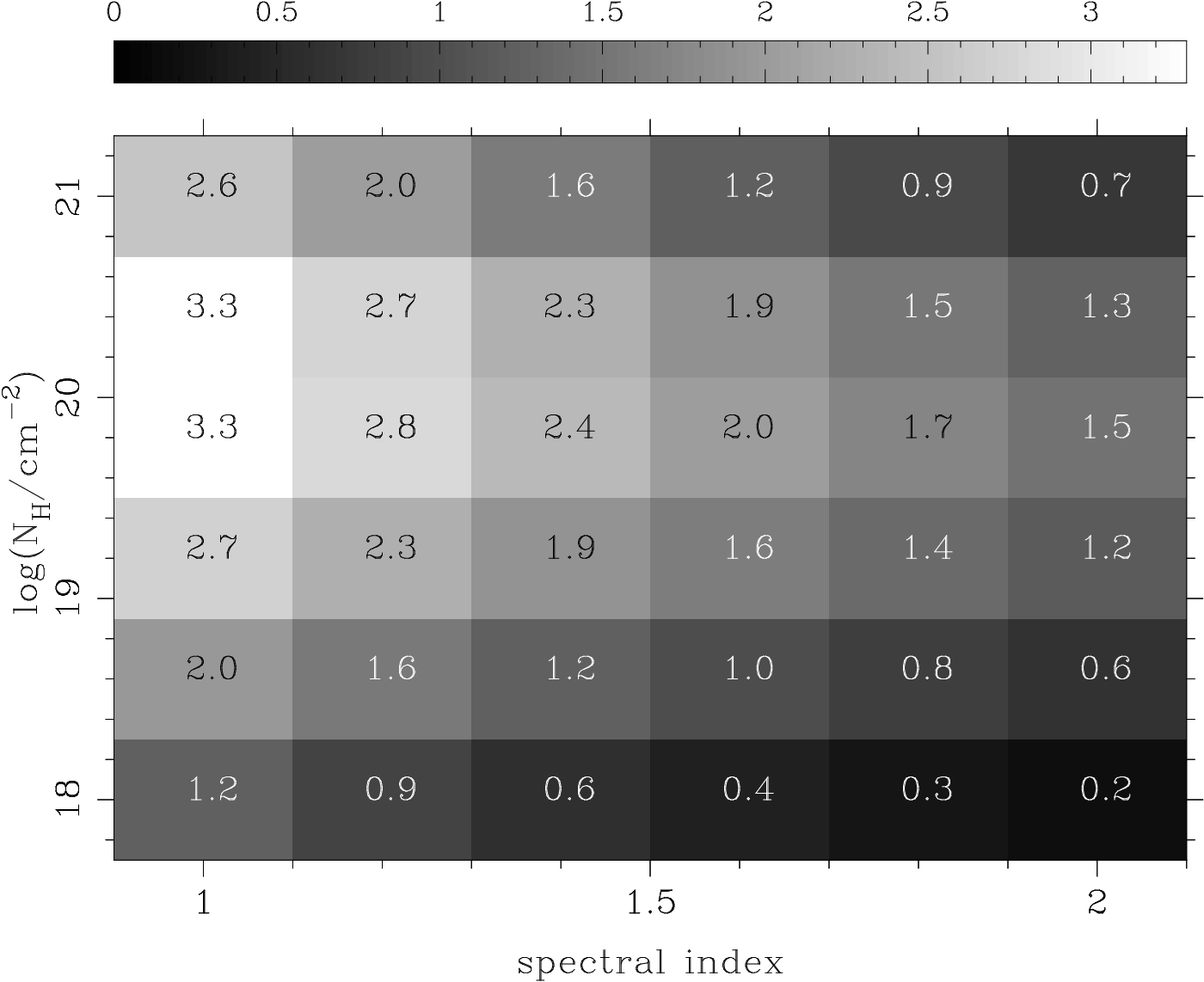}
\includegraphics[height=3in,angle=270]{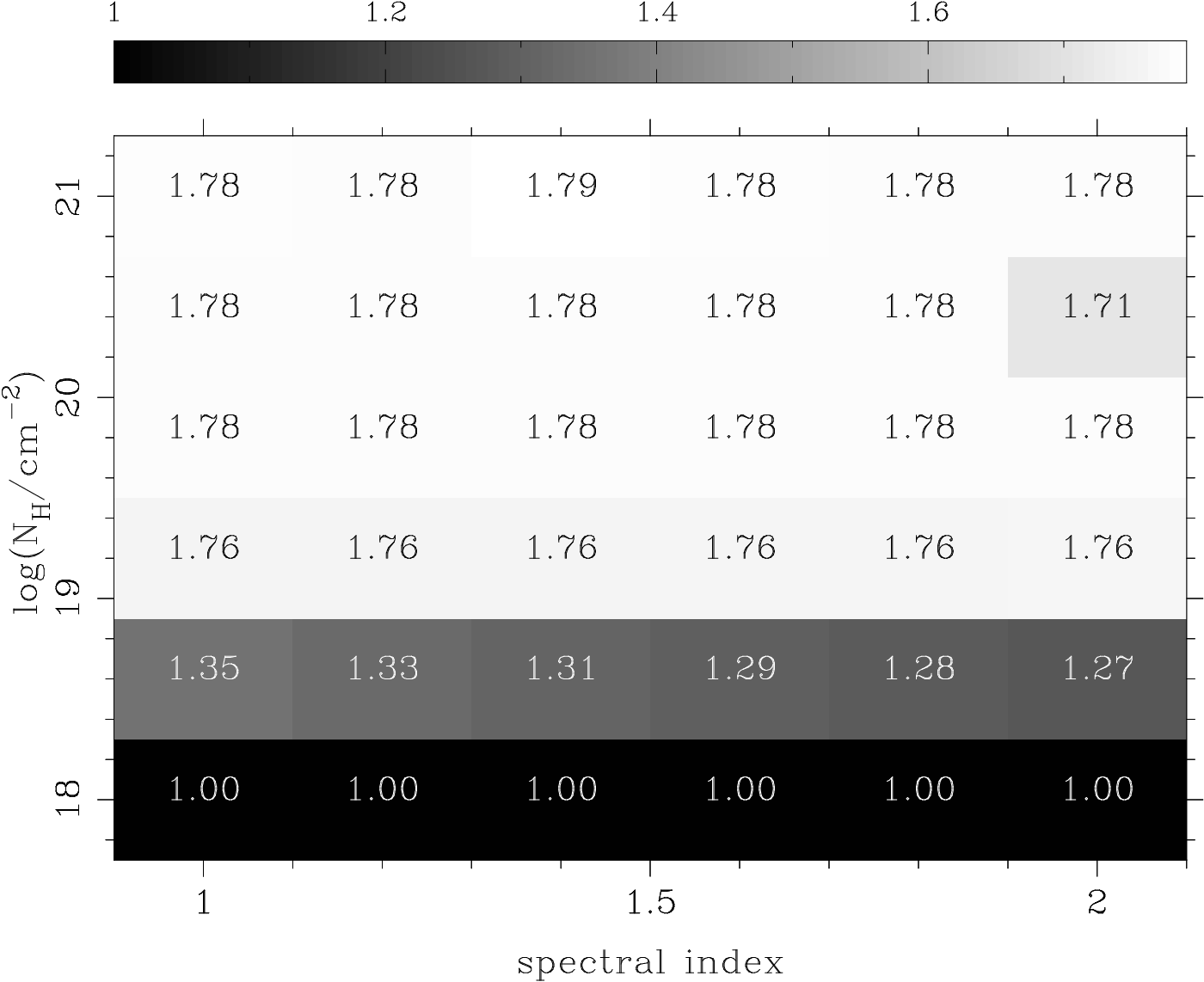}
\caption{Widths with $C=5$ at $3.16 \times 10^7$ years as measured by
  $d_{0.1}$ (in Mpc, top) and $r_{-2.5}/r_{-3}$ (bottom).}\label{fC5}
\end{figure}

Figure \ref{fC5} reproduces the grids of spectral parameters with a
clumping factor of $C=5$. The thickness measures follow the same
trends seen with $C=1$. Increasing the clumping factor produces almost
the same grid of $d_{0.1}$ values because it has little effect on the
high-energy photons that pre-ionized the outer region. 

The $r_{-2.5}/r_{-3}$ ratios on the $C=5$ grid are quite a bit higher
than with $C=1$. The ratio exceeds $1.2$ for all spectra with
$N_\mathrm{H} \ge 10^{18.6} \: \mathrm{cm}^{-2}$. A clumping factor of
$C = 20$ (also not shown) produces a very similar grid for the more
obscured spectra, and produces even higher ratios for $\logNH \le 19.2$.

\subsubsection{The Ionizing Background}\label{sssBG}

The IGM is expected to have already undergone some non--negligible
ionization by galaxies, and possibly by fainter quasars, by the time
the massive black holes producing the observable $z\ga 6$ quasars
appear. The ionizing background may have both a smooth X-ray component
\citep[e.g.][]{Venkatesan_etal2001, Oh2001} from pre--existing fainter
quasars, and a ``swiss-cheese''-like component of smaller ionized
bubbles surrounding (clusters of) galaxies, the latter of which may be
highly clustered around the massive halos hosting our quasars of
interest \citep[e.g.][]{WL2007,LMZ2007,AA2007}.

\begin{figure}
\includegraphics[height=3in,angle=270]{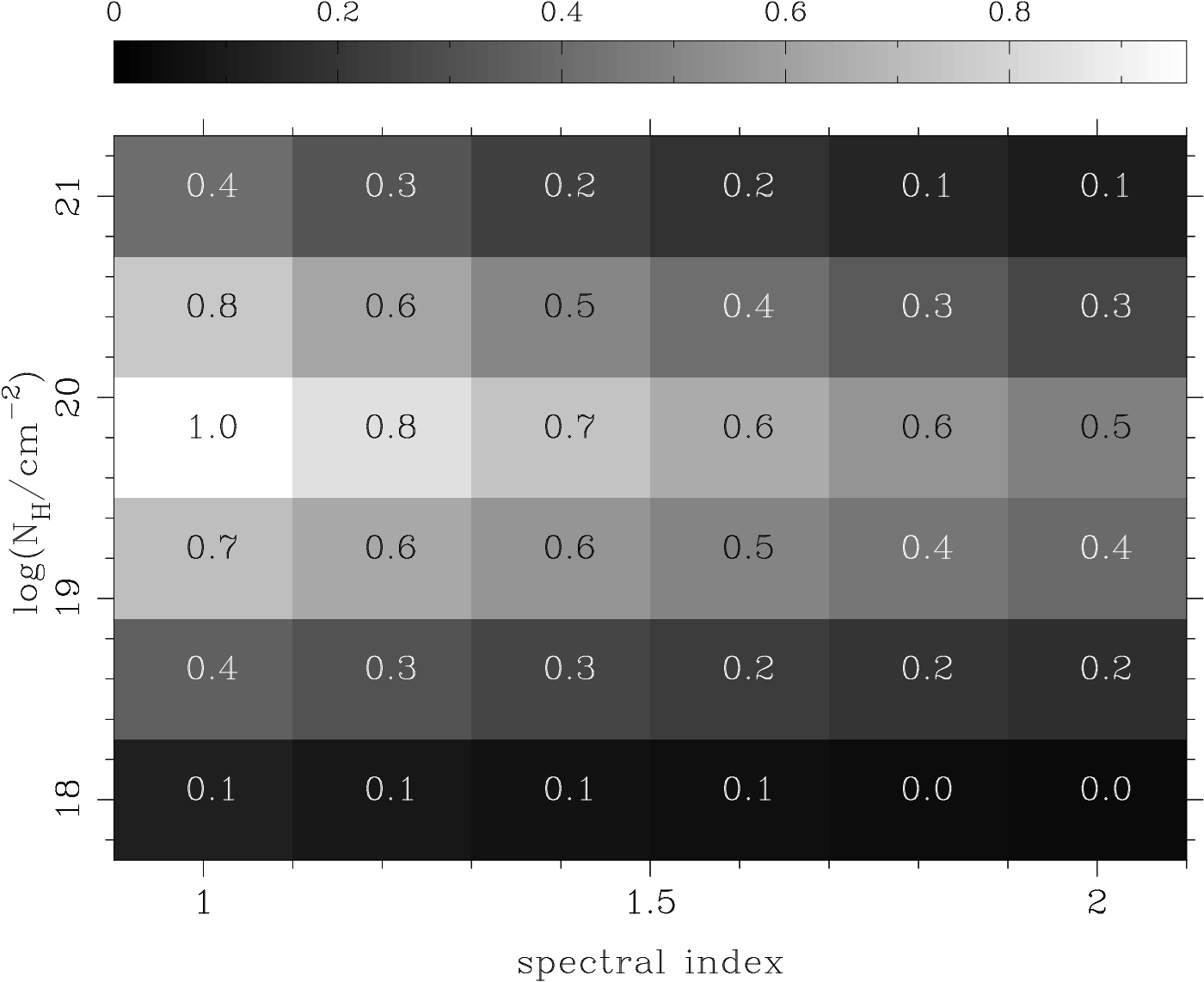}
\includegraphics[height=3in,angle=270]{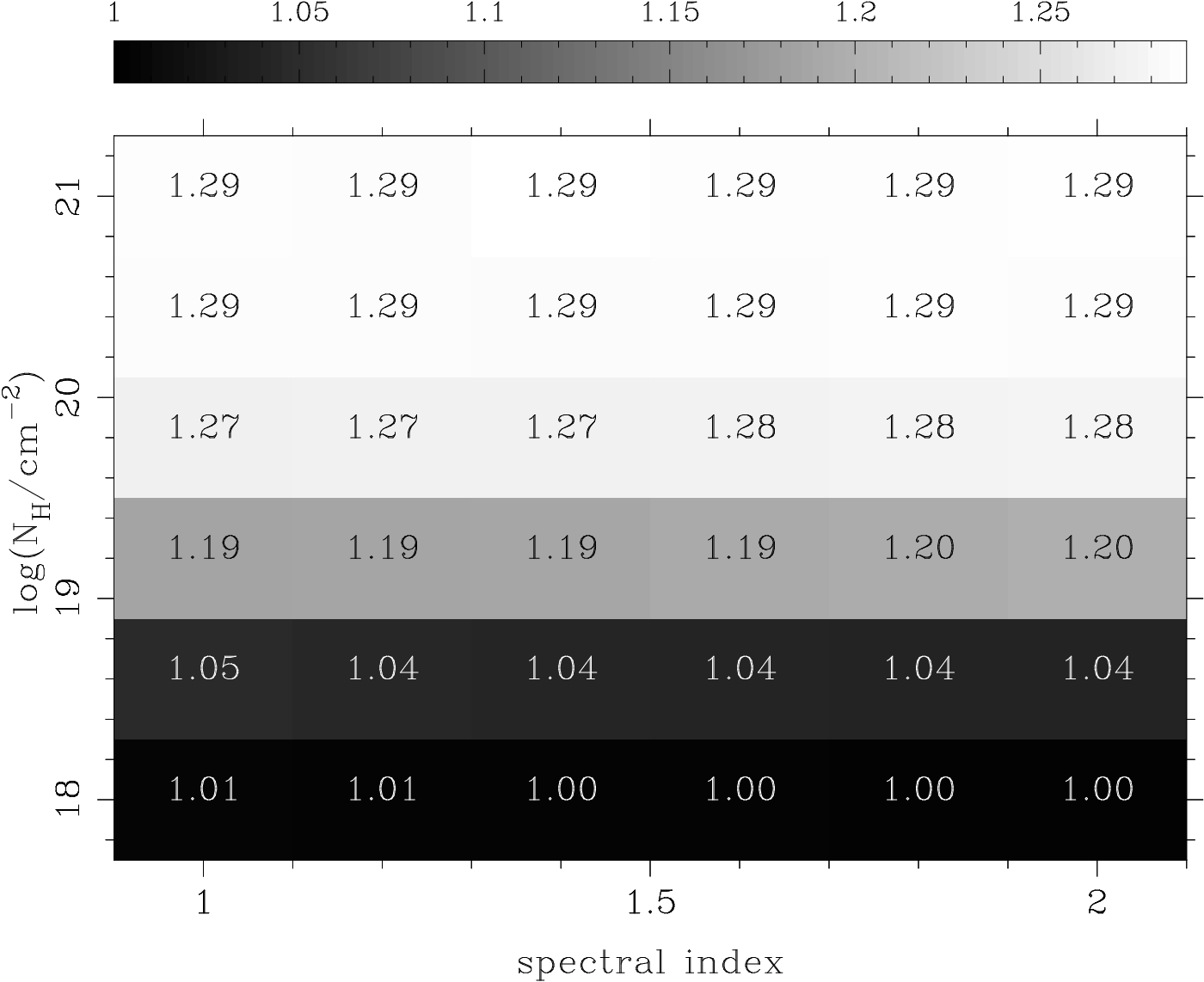}
\caption{Widths with a uniform ionizing background, $X_{\Hp, BG} =
  0.75$. Note the here we are using the distance from $X_{\Hp} = 0.9$
  to $X_{\Hp} = 0.8$ to measure the outer front (top) and still using
  $r_{-2.5}/r_{-3}$ (bottom) for the inner front. Both are at $3.16
  \times 10^7$ years.}\label{fBG025}
\end{figure}

Here we first calculate a grid of thickness measures with a purely
uniform background $\Gamma_{BG} = 4.4 \times 10^{-17} \:
\mathrm{s}^{-1}$ (see the discussion of radiation transfer below),
producing an equilibrium $X_{\Hp, BG} = 0.75$.  We will then study
three variations of this model, to address the following effects: (i)
modification of the background flux due to radiative transfer effects
caused by the ionization by the quasar, (ii) a non--uniform
(swiss--cheese) background ionization topology, and (iii) enhanced
ionization near the quasar due to clustering of pre--existing
galaxies.

In the presence of a background, there are fewer neutral hydrogen
atoms to ionize, and the ionization fronts travel farther in a given
amount of time. The lower neutral hydrogen density also means the
m.f.p. is longer, so the fronts are somewhat thicker. The equilibrium
neutral fraction is relatively unaffected as long as it is well below
$1 - X_{\Hp, BG}$.  Setting $X_{\Hp, BG} = 0.75$ obviously makes our
previously--defined measure for the thickness, $d_{0.1}$, unusable, so
we here instead examine the distance from $X_{\Hp} = 0.9$ to $X_{\Hp}
= 0.8$ (and we can still use $r_{-2.5}/r_{-3}$). Both of these
thickness--measures are shown in Figure \ref{fBG025}, again at $3.16
\times 10^7$ years, as a function of spectral slope and obscuring
column density.

The new measure of the outer part of the front behaves much like
$d_{0.1}$ did for the grids with no background (except that it is
obviously smaller overall). The figure also shows that
$r_{-2.5}/r_{-3}$ is somewhat larger than for the no--background
case. The reduced neutral hydrogen density allows the front to move
more quickly, but has little affect on the equilibrium ionized
fraction where $X_{\Hn} \ll 1 - X_{\Hp, BG}$. This means that the
inner (most highly ionized) edge of the front slows as it approaches
the equilibrium value, while farther out the front is still receding
at a faster rate, effectively thickening the front.

There is an inconsistency in the way the background ionization rate
was added to the quasar flux above, however (which is shared by many
other similar studies in the literature). We specified an ionization
rate per neutral hydrogen atom $\Gamma_{BG}$ which is constant,
equivalent to a uniform background flux that does not depend on the
ionization state of the gas. In fact, the flux from a homogeneous
population of emitters will depend on the mean free path, and hence on
the ionization state of the gas. The more highly ionized the gas is,
the farther ionizing photons will travel through it, meaning that more
distant sources can contribute to the local flux. In fact, the mean
free path within a highly ionized \ion{H}{ii} region might exceed the
size of the region, meaning that the background will be the sum of the
flux from all the sources inside the region, and the background will
grow as the region expands.

In a homogeneous medium with a total photon emission coefficient
of $j$ (photons per unit time per unit volume), the total
incident flux (integrated over all angles) is simply\footnote{Usually
an emission coefficient is defined as the amount of power emitted per
unit solid angle per unit volume. Here we use a somewhat different
definition which is convenient for the task at hand.}
\begin{equation}\label{RTuniform}
J = \frac{j}{\nH (1 - X_{\Hp}) \sigma} = j \, l_\mathrm{MFP}
\end{equation}
where $\sigma$ is the absorption cross section for the photons. Since
$\Gamma = J \sigma$, the background ionization rate should clearly
depend on the ionization state of the gas. Similarly, the flux in the
center of a sphere of radius $r$ (where $l_\mathrm{MFP} \gg r$, and $j =
0$ outside the sphere) is $J = j \, r$.

In our model the medium is obviously not homogeneous, since $X_{\Hp}$
varies with position and time (and later we will be interested in a
variable $j$ as well). Solving for the background flux $J$ exactly in
an inhomogeneous medium requires computationally expensive 3-D
radiation transfer (and still 2-D in the azimuthally symmetric case), 
so we have implemented an approximate radiative
transfer algorithm to estimate the effect using a 1-D calculation. 
We assume all of the
background photons are at $13.6$ eV and calculate the flux using the
equation:
\begin{equation}\label{RTapprox}
J(r) = J_0(r) + J_\mathrm{in}(r) + J_\mathrm{out}(r)
\end{equation} 
\begin{equation}
J_0(r) \equiv j(r) \, l_\mathrm{MFP}(r) \, [1 -
\exp(- \Delta r / l_\mathrm{MFP}(r))]
\end{equation} 
\begin{equation}
J_\mathrm{in}(r) \equiv \frac{1}{2} \int_0^{r-\Delta r} j(r')  \left(\frac{r'}{r}\right)^2
e^{[\tau_0(r') - \tau_0(r)]} dr' 
\end{equation}
\begin{equation}
J_\mathrm{out}(r) \equiv \frac{1}{2} \int_{r+\Delta r}^{\infty} j(r') e^{[\tau_0(r) - \tau_0(r')]} dr'
\end{equation}
where the first term is the flux from within one
radial step $\Delta r$, the second term is the estimated flux from
within $r - \Delta r$, the third term is the estimate flux from
outside of $r + \Delta r$, and $\tau_0(r)$ is the optical depth at
$13.6$ eV as defined by equation \ref{optical_depth_intergral}. The
\textit{ad hoc} factor $\left(\frac{r'}{r}\right)^2$ is used to
decrease the contribution to the flux from spherical shells at small
$r$ (roughly approximating the decrease in the total emitting volume
in 3D at small radii). At $r \gg l_\mathrm{MFP}$ with uniform $j$ and
$X_{\Hp}$, equation \ref{RTapprox} reproduces the analytical result
(equation \ref{RTuniform}).

\begin{figure}
\includegraphics[height=3in,angle=270]{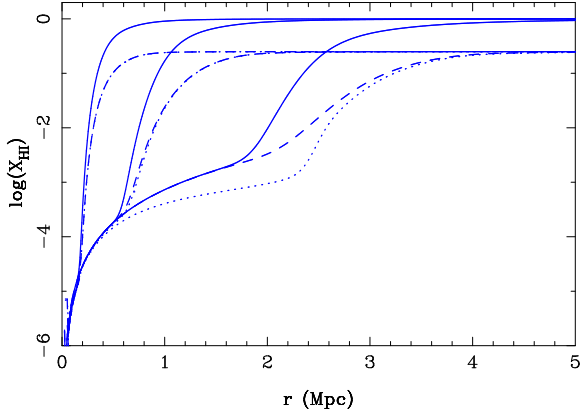}
\caption{Comparison of neutral fraction with no ionizing background
  (solid), with a uniform background without background radiation
  transfer (dashed), and the same background with approximate RT
  (dotted) at $\log(t/\mathrm{years}) = 6$, $7$, and $8$ (left to
  right). Spectral parameters are the fiducial ones used in Figure
  \ref{fFrontEvolution} and the backgrounds are chosen to produce
  $X_{\Hp, BG} = 0.75$. Note that the RT effects make a noticeable
  difference in the $10^8$ year case: the reduction of the mean free
  path by the quasar boosts the contribution of background galaxies to
  the total ionization rate inside the \ion{H}{ii} region.}
  \label{fBGCompare}
\end{figure}

Figure \ref{fBGCompare} compares the ionization structures with and
without the approximate RT algorithm. The photon emission coefficient
is $j = \alpha_B \, n_\mathrm{H}^2 \, X_{\Hp, BG}^2= 8.3 \times
10^{-22} \: \mathrm{s}^{-1} \mathrm{cm}^{-3}$. For quasar ages of
$\la 10^7$ years, the radiation transfer makes no difference to
the ionization structure. The central ionized region is simply too
small to allow a large additional background to build up. At $10^8$
years, RT decreases the neutral fraction, but only for $-4 \la
\log(X_{\Hn}) \la -1.5$. This is because the ionized region has
grown large enough for the diffuse background to become significant,
but the quasar flux still dominate close the quasar, and the MFP is
still very small for $13.6$ eV photons once the neutral fraction gets
large, so the extra flux is confined to the inner part of the
front. We conclude that radiation transfer of flux from a uniform
ionizing background could be important for Lyman-series observations
of old quasars or with high backgrounds, but will be less important
for 21 cm observations.

\begin{figure}
\includegraphics[height=3in,angle=270]{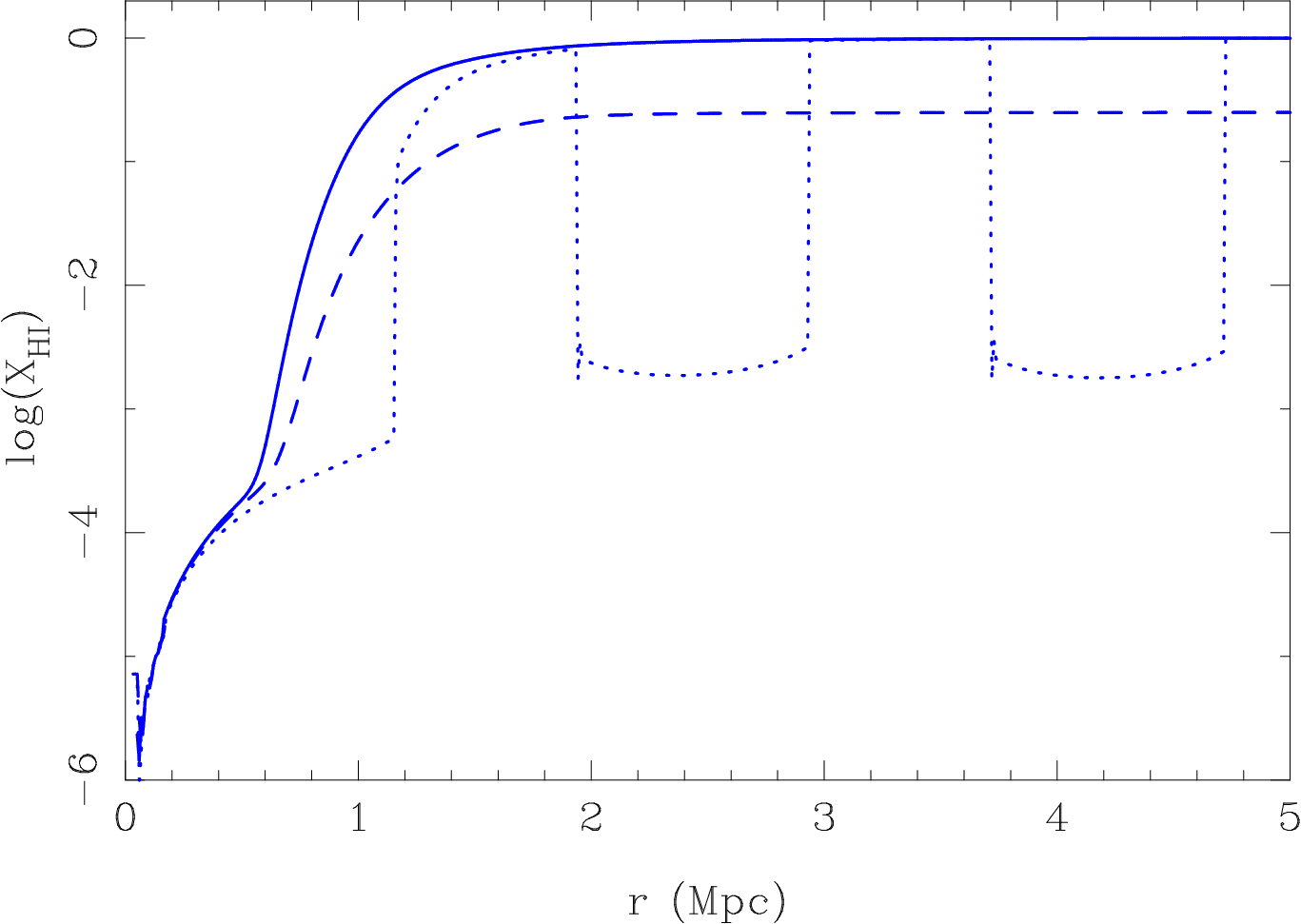}
\includegraphics[height=3in,angle=270]{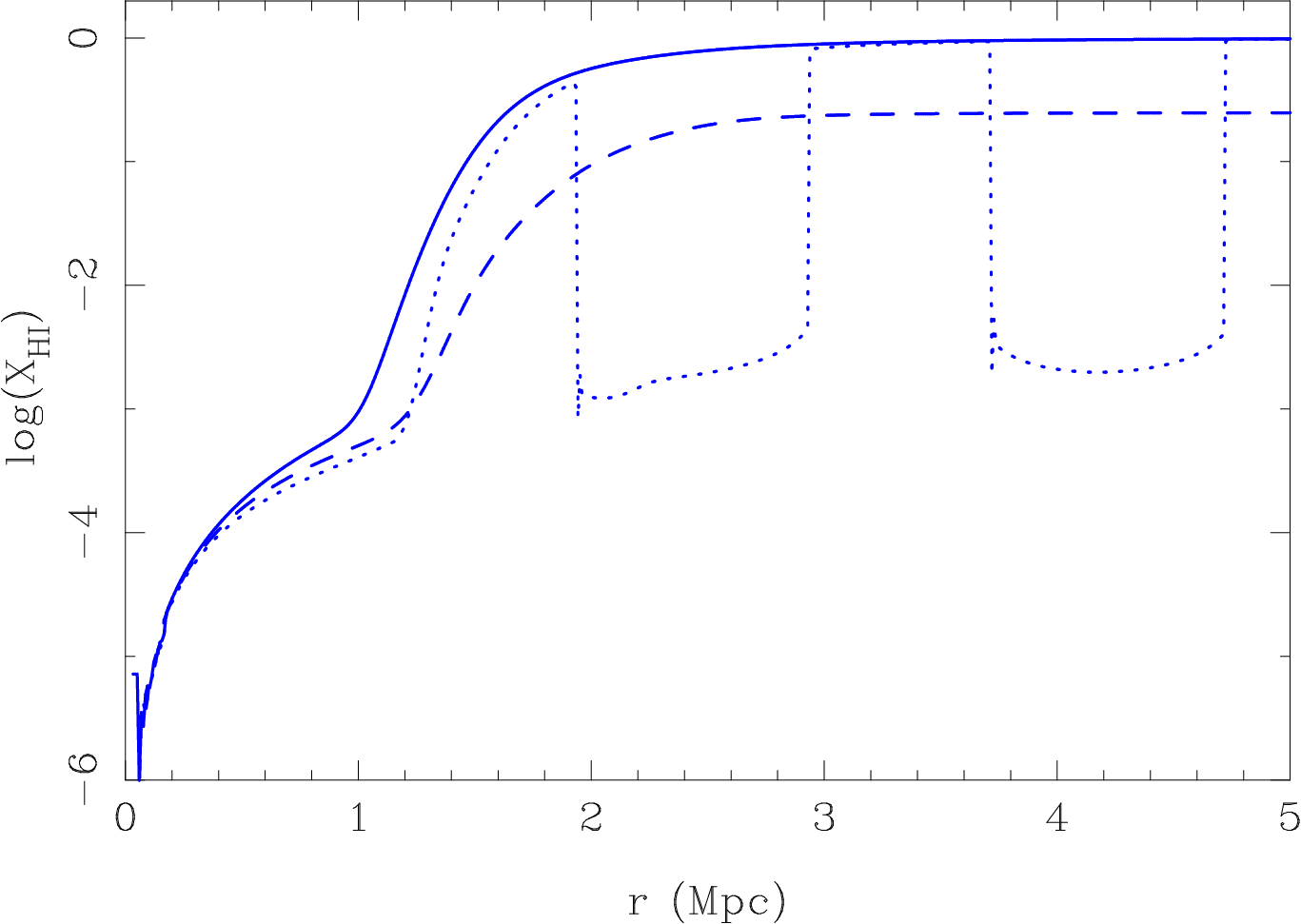}
\includegraphics[height=3in,angle=270]{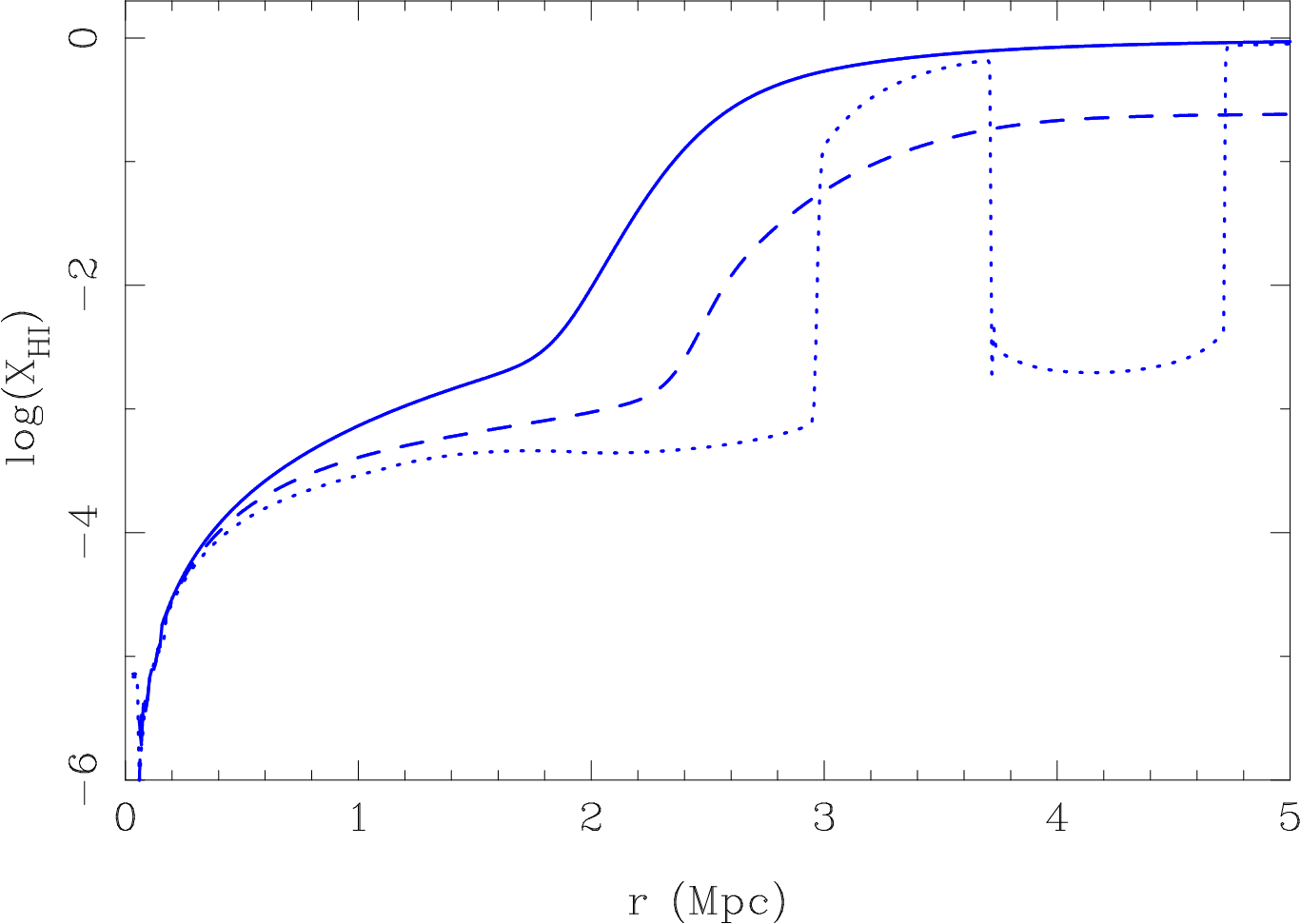}
\caption{Comparison of the neutral fraction with no ionizing background
  (solid), with a smooth ionizing background (and approximate radiation
  transfer) chosen to produce $X_{\Hp, BG} = 0.75$ (dashed) and with a
  ``swiss-cheese'' background (dotted) of $1$ Mpc ionized bubbles
  separated by $0.79$ Mpc neutral regions, which would produce a similar
  average ionization. Spectral parameters are the
  fiducial ones used in Figure \ref{fFrontEvolution}. The panels show
  the ionization structures at $10^7$, $3 \times 10^7$, and $10^8$
  years, respectively.}
  \label{fPFCompare}
\end{figure}

Next, let us examine the effect of a patchwork of pre-existing ionized
bubbles along our line of sight. This is the type of background
profile that results from the ``swiss-cheese'' topology of small
ionized bubbles (generated by neighboring galaxies and possibly AGN)
embedded in a neutral IGM. Naively, we might choose to fill $75\%$ of
the IGM with ionized bubbles in order to compare this scenario with
the smooth background of Figure \ref{fBGCompare}, but here we have to
be careful about what quantities we want to compare. The recombination
rate in a fully ionized region is $\alpha_B \, n_\mathrm{H}^2 = 1.48
\times 10^{-21} \: \mathrm{s}^{-1} \mathrm{cm}^{-3}$, so if a fraction
$f_\mathrm{ion}$ of the universe is contained in ionized bubbles, then
the average photon emission coefficient needed to maintain equilibrium
is $j = \alpha_B \, n_\mathrm{H}^2 f_\mathrm{ion}$. Equating this to
the coefficient for the smooth $X_{\Hp, BG} = 0.75$ case we obtain
$f_\mathrm{ion} = 0.75^2 = 0.56$. In fact we will need an even higher
background in the bubbles in order for them to be highly ionized, as
we will discuss below.

Next we need to determine the size of the ionized
bubbles. \citet{FZH2004} calculated that when $56\%$ of the IGM was
contained in galactic \ion{H}{ii} regions, the \ion{H}{ii} region radius
distribution peaked at $\sim 4$ Mpc comoving, or about $0.6$ physical
Mpc at $z=6$.  We have set up a background ionization profile
consisting of evenly-spaced ionized regions with a diameter of $1$ Mpc,
occupying $56\%$ of the IGM volume (and therefore separated by neutral
regions of $0.79$ Mpc along the LOS).  The neutral fraction within the
bubbles is initially set to $10^{-3}$.  The background needed to
maintain this level of ionization is
\begin{equation}
j = \frac{\alpha_B \, n_\mathrm{H} \, X_{\Hp, BG}^2}
{\sigma_0 \, r \, (1-X_{\Hp, BG})} = 2.01 \times 10^{-21} \: \mathrm{s}^{-1} \mathrm{cm}^{-3} 
\end{equation}
where $r = 0.5 \: \mathrm{Mpc}$ is the radius of the bubble. This is
$35\%$ higher than the background in the smooth case. We again use
equation \ref{RTapprox} to approximate radiation transfer for the
background photons.

The resulting ionization structure is shown in Figure
\ref{fPFCompare}. Without the background, the fronts in the
inter-bubble regions would be essentially the same shape as they are
in totally neutral IGM, except sliced and shifted apart where the
bubbles occur, since the ionizing radiation traverses the bubbles
unabsorbed (except by helium and the small amounts of residual
hydrogen) so the front can pick up where it left off on the other
side. The presence of helium and the geometric dilution cause some
changes in the shape, but the background radiation transfer has a
greater effect. Once the front has swept past a bubble, it's
background photons can escape, and they stream out to contribute to
the ionization of the inter-bubble regions. 

This type of background structure may have interesting observational
manifestations. Within the inter-bubble regions, the front is somewhat thinner
than with either no background or a smooth background. The front is
thinner than in the no-background case because the background photons
contributing to the movement of the front have a short MFP. The front
is thinner than in the smooth-background case because it is
``trapped'' by the fully neutral inter-bubble regions where  
the mean free path is short, so it is slowed down,
meaning that a more highly ionized part of the front (e.g. $r_{-3}$)
can ``keep up'' better with a more neutral part (e.g. $r_{-2.5}$),
resulting in a thinner front over all. There will be brief windows
when $r_{-3}$ is still trapped in the residual neutral gas on one side
of a bubble, while $r_{-2.5}$ is on the far side, but it would require
detailed simulations of absorption spectra to determine the
observability of such a situation. The shape of the front at neutral
fractions more relevant to 21 cm observations again tends to suggest
smaller thickness measurements, but it seems likely that the
thicknesses could still be measurable.

The signature of the hard quasar spectrum is also still quite visible
on large scales, for instance in the separation between $X_{\Hn} \sim
10^{-4}$ and $X_{\Hn} \sim 0.1$. This suggests that a combination of
21 cm and Lyman-series observations, particularly when smoothed on
large scales or averaged over many quasars, could place constraints on
the ionizing SED of the quasar or quasars.  If the patchy ionization
structure is resolved in future, high--resolution 21cm observations
with SKA, then the thickness of the front in--between the individual
galaxy--bubbles can reveal evidence for a hard quasar spectrum separately
along each direction measured from the quasar.

\begin{figure}
\includegraphics[height=3in,angle=270]{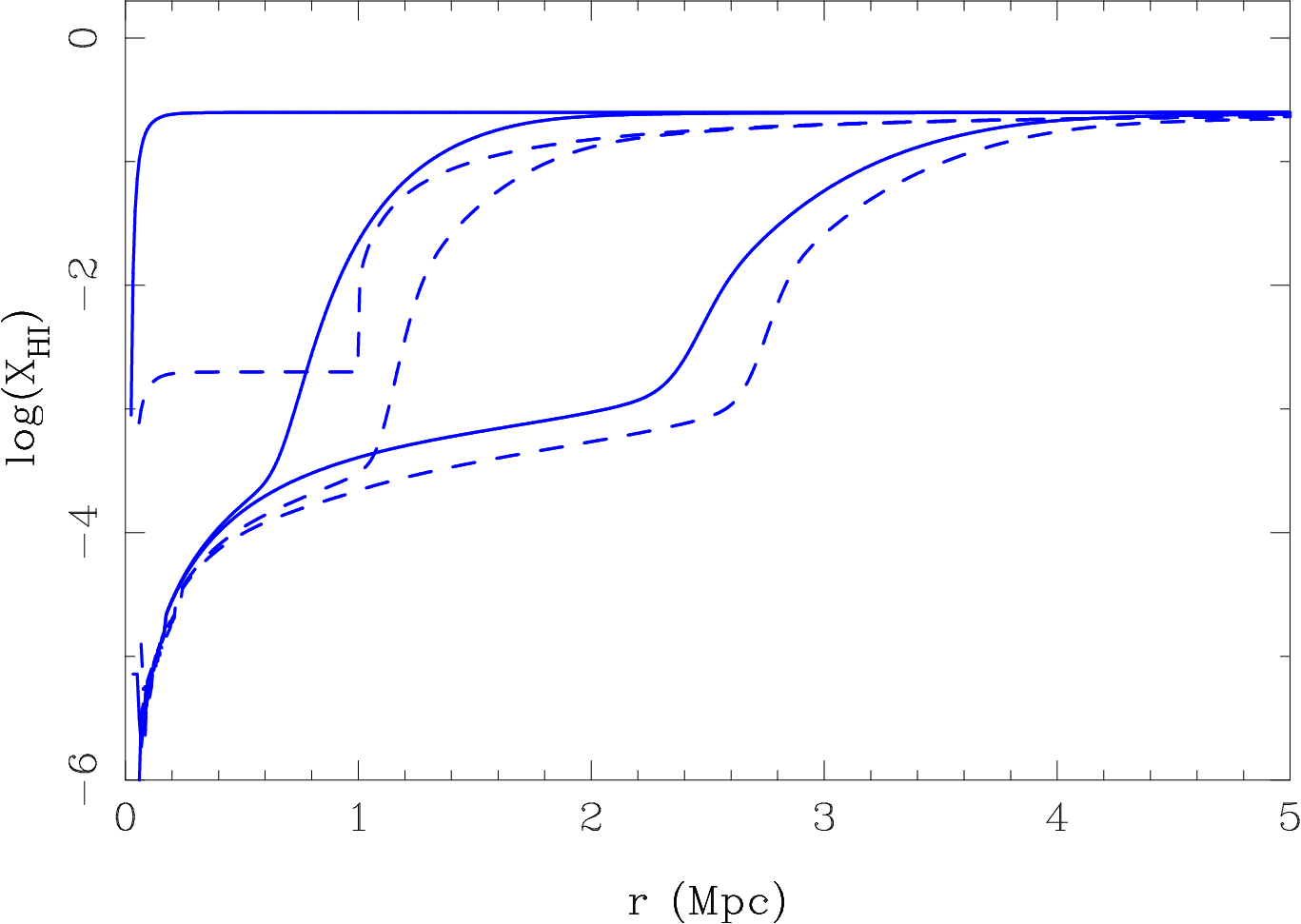}\\
\includegraphics[height=3in,angle=270]{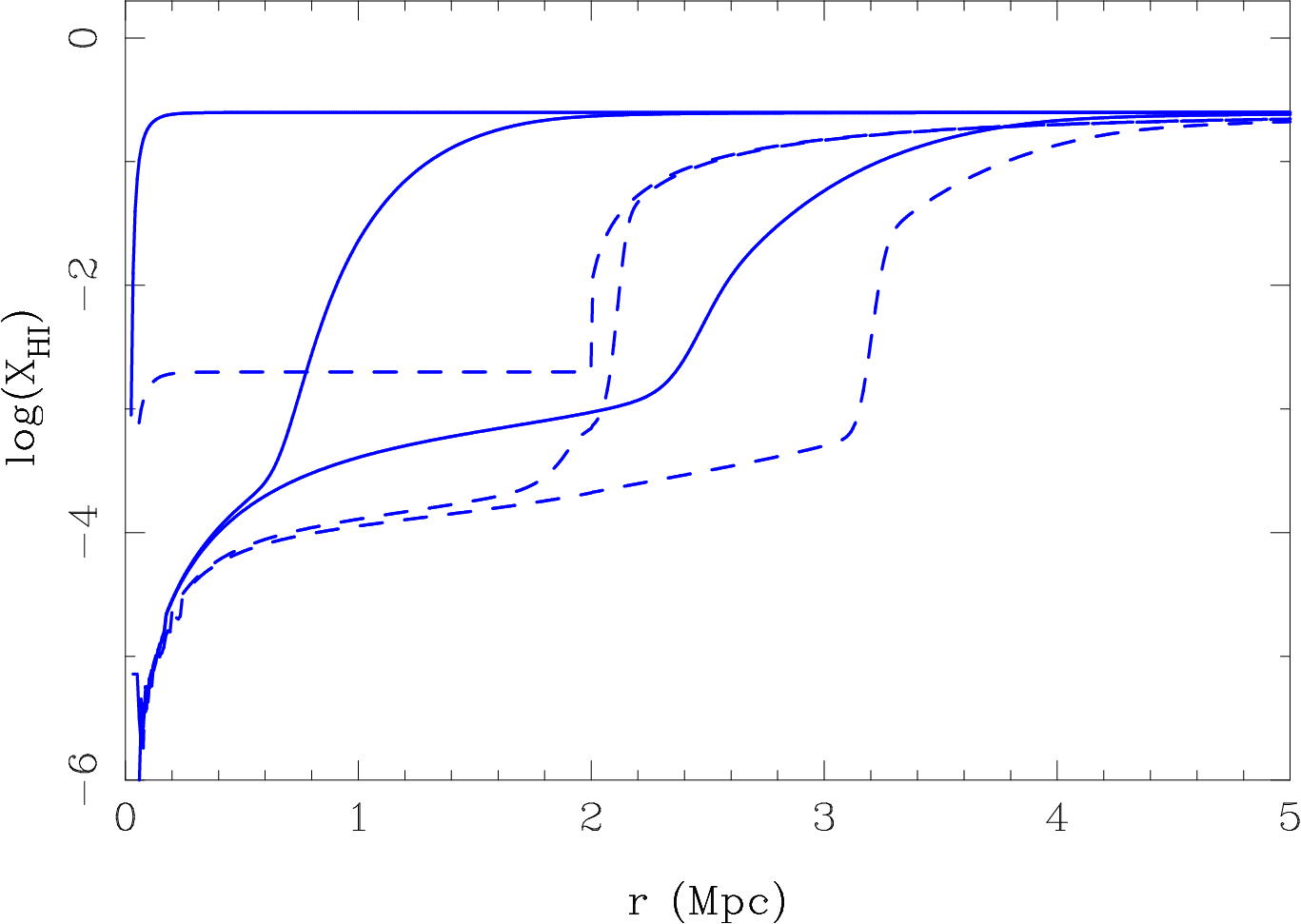}
\caption{Comparison of the neutral fraction with a uniform ionizing
  background (solid) and a biased background from pre--existing
  galaxies clustered around the quasar (dashed). Approximate
  background RT is used in both cases. Spectral parameters are the
  fiducial ones used in Figure \ref{fFrontEvolution}. The three sets
  of curves correspond to source ages of $\log(t/\mathrm{years}) = 4$,
  $7$ and $8$, respectively. In the first panel, the central
  pre-ionized bubble has a radius of $1$ Mpc, while in the second
  panel the radius is $2$ Mpc.}
  \label{fFrontWL}
\end{figure}

Finally, we consider the effect of clustering of background
sources around the quasar. \citet{WL2007} have explored the mean
ionization structure around a quasar resulting from the biased
clustering of galaxies around its host halo (not addressing the
discreteness of the bubbles in the radial profile).  They point out
that even outside the I-front, the IGM is ionized 10-20\% more than
the global mean ionized fraction.  While a high--resolution 21cm
interferometer, such as SKA, could resolve such clusters of
galaxy--bubbles outside the quasar's \ion{H}{ii} region, the next
generation of experiments will likely not have sufficient
resolution. In this case, the I--front will appear extended due to the
presence of galaxies.

In order to explore the effect of such a pre--ionization profile on
our conclusions, we implemented an ionizing background designed to
qualitatively mimic the background ionization structure predicted by
\citet{WL2007} when the IGM far from the quasar is at $X_{\Hp, BG}
\sim 0.7$--$0.8$. In this scenario pre--existing galaxies produce an
\ion{H}{ii} region centered on the quasar; outside this \ion{H}{ii}
region, there is a tail of excess partial ionization, slowly
approaching the background level of the neutral fraction in the IGM.

To produce the first panel in Figure \ref{fFrontWL}, we set up an
ionizing background to produce a $1$ Mpc radius inner ionized region
(initially set at $X_{\Hp} = 2 \times 10^{-3}$ with $j = 4.41 \times
10^{-21} \: \mathrm{s}^{-1} \mathrm{cm}^{-3}$). Outside of that, the
profile is set to
\begin{equation}
 X_{\Hp, BG} = 0.75 + \left[\frac{(r/\mathrm{Mpc})  + 1.91}{1.81}\right]^{-3},
\end{equation}
with $j = \alpha_B \, n_\mathrm{H}^2 \, X_{\Hp, BG}^2$ for $r > 1 \:
\mathrm{Mpc}$.  In the second panel the radius of the pre-ionized region
is increased to 2 Mpc and the outer profile is shifted accordingly.

Figure \ref{fFrontWL} illustrates what happens to the I--front as it
passes through such a biased background. The dashed curves show the
I--front expanding into the pre--ionized region, contrasted with the
case when the pre--ionization is uniform (solid curves).  The figure
reveals that once the front has passed beyond the high-background
inner region, it quickly begins to approach the shape of the
ionization profile in a uniform background, though the extra flux from
the sources clustered close to the quasar keeps it ahead of the front
in the uniform case. The presence of helium is particularly important
in this case because it determines the speed at which the front can
propagate through the pre-ionized region (since helium is assumed to
be unaffected by the background and therefore still neutral). We have actually
chosen rather small sizes for our central pre-ionized
region. \citet{WL2007} predict central regions of $r \sim 2$--$8$ Mpc
when $X_{\Hp, BG} \sim 0.7$--$0.8$. A larger region implies a larger
background flux once the front has passed the inner region. This would
prolong the time during which the front shape is affected by the
background bias. Even a region as small as $2$ Mpc results in fronts
whose inner parts are significantly less broad at $10^8$ years, as seen in the second
panel.

\section{Caveats}\label{sCaveats}

Before we summarize our conclusions, let us briefly review the
limitations of the methods we have employed in reaching them. 

The one-the-spot approximation --- in which we assume that ionizing
photons produced by recombination are absorbed at their point of
origin --- is strictly valid only where the mean free path of the
recombination photons is much shorter than other length scales of
interest, such as the thickness of the front.\footnote{The situation is
slightly different for soft (e.g. stellar) spectra, in which the mean
ionizing photon from the source is also close to the threshold. In
that case both the source spectrum and the recombination radiation are
concentrated near the threshold. See \citet{OsterbrockAndFerland}.}
This condition is violated where the neutral fraction is very low, in
the inner part of the ionized region. It is also less accurate for
helium-derived photons, since they are generally more energetic. The
result is a slight overestimate of the ionization rates, and therefore
a slight underestimate of the neutral fraction, close to the
quasar. The effect is less important for harder spectra (because they
leave higher neutral fractions overall), and is unimportant in the
outer part of the front. For instance, the mean free path at $13.6 \:
\mathrm{eV}$, $l_\mathrm{MFP}(\nu_0) < 0.01 \: \mathrm{Mpc}$ for
$X_{\Hn} > 0.07$.

We have ignored the propagation time for photons (due to the finite
speed of light). Our results accurately predict the observed structure
along a line of sight aimed directly at the quasar, in an absorption
spectrum, for example. To simulate two dimensional projections of the
ionization structure (like 21 cm maps), we would need to apply a
simple transformation (see citations in \S~\ref{ssIonization}). In
fact, since the thickness of the I--front varies with time, in
principle, a comparison of the thickness in transverse and radial
directions, in addition to the angular dependence of the overall
bubble size due to finite--speed--of light effects
\citep{WLB2005,Yu2005,SIA2006}, will contain information about the age
of the quasar.

We have assumed the density of the IGM surrounding the quasar is
homogeneous on a macroscopic scale, with smaller variations accounted
from by a uniform clumping factor. Quasars are actually expected to
form in biased environmental conditions, with high gas density and a
high density of neighboring halos. The largest effect would be along
lines of sight that intersect high-density Lyman limit systems,
which could stall the propagation of the I--front and shield the gas
further along the LOS from ionizing photons. The less extreme effects
of density inhomogeneities would be somewhat ameliorated by certain
observational necessities. In order to get a large enough signal to
noise ratio to meaningfully constrain the structure of quasar
\ion{H}{ii} regions, it may be necessary to stack 21 cm images of
multiple bubbles. Similarly, measurements of the spectra of many
quasars would be combined in order to constrain the ionization
structure with Lyman-series absorption observations. In either case,
density inhomogeneities would add to the random ``noise'', but could,
to some extent, be averaged out. Width measurements may be more robust
under such conditions than the simple size measurements discussed by
\citet{Maselli_etal2007}, however more study is needed to determine
that conclusively.

As mentioned in section \ref{ssPhyicalParameters}, we have assumed the
IGM is isothermal, justifying this by pointing out that most of the
heating will occur before recombinations become important. Therefore,
as long as we choose the gas temperature to match conditions expected
inside the \ion{H}{ii} region, this assumption should have little
effect on our results. 

Finally, we should remind the reader that since we did not actually
simulate absorption spectra, we have not taken into account the effect
of the GP trough damping wing (from the neutral hydrogen surrounding
the ionized region), which could be very important for the observational
determination of $R_\beta/R_\alpha$. In fact, \citet{MH2004}
essentially use this effect to constrain the neutral hydrogen fraction
in the IGM.

\section{Discussion and Conclusions}\label{sConclusions}

We have found that high-redshift quasars residing in a partially
neutral IGM could produce ionization fronts with an observable
thickness, given a sufficiently hard ionizing spectrum. We discovered that
simulating the time-dependent evolution of the front is crucial for
making accurate predictions of the ionization structure, and that the
presence of helium in the IGM and secondary ionizations by high-energy
photoelectrons can both have significant effects on the ionization
structure. 

With an intrinsic hydrogen column density $\logNH \ga 19.2$ or a
sufficiently hard power--law spectrum combined with some obscuration
(e.g. spectral index $s \la 1.2$ at $\logNH \ga 18.0$), the outer
thickness of the front exceeds $\sim 1$ physical Mpc and may be
measurable from the three--dimensional morphology of its redshifted
21cm signal.

The highly ionized inner part of the front, which may be probed by
Lyman line absorption spectra, remains thin for bright quasars unless
a large obscuring column ($\logNH \ga 19.2$) removes most of their
ionizing photons up to $\approx 40 \: \mathrm{eV}$. Otherwise the
lowest energy photons with the shortest mean free paths always
dominate ionization at the inner face of the front resulting in a
small thickness. 

Highly absorbed ionizing spectra leave a relatively large neutral
fraction within the \ion{H}{ii} region, which means that the
Lyman-series optical depths can be large even within the front. For
sources with $\logNH \ga 19.8$, the Lyman $\alpha$ trough (where the
neutral fraction is $\ga 10^{-3}$) underestimates the size of the
\ion{H}{ii} region by a factor of $\ga 4$. The bias can get as large
as a factor of $\approx 8$ within out parameter space. This is in
addition to other effects that bias this measurement by a much smaller
amount, as already discussed by several authors
\citep{Maselli_etal2007,LMZ2007,BH2007b}.

These obscured spectra also result in a large difference between the
sizes measured by the Lyman $\alpha$ trough, and those measured by the
Lyman $\beta$ trough (where the neutral fraction is $\ga
10^{-2.5}$). We estimate that $R_\beta/R_\alpha$ could be higher than
$1.2$ for $\logNH \ga 19.8$ with a quasar age of $3 \times 10^7$
years.

We explored the effects of uniform and non-uniform ionizing
backgrounds, finding that even with a large uniform background the
thickness of the ionization front has the potential to constrain the
spectral parameters of the quasar. An ionization front propagating
into a ``swiss-cheese'' background of small pre-ionized bubbles
obviously makes for a more complicated ionization structure, but
again, the shape of the front on both large and small scales is
controlled largely by the quasar's ionizing SED. The greatest
challenge is presented by a quasar turning on within a large
pre-existing ionized region due to galaxies clustered around the
quasar. In this case, the thickness of the front
does not reflect the quasar's spectrum until the quasar has pushed the
front a significant distance farther into the IGM. A measurement of a
thin front around a quasar cannot, therefore, be interpreted as
definitive evidence of a soft or unabsorbed quasar spectrum. On the
other hand, measuring a thick front does suggest a highly absorbed
quasar spectrum, because the galactic sources producing the pre-quasar
ionization should be soft.

\citet{SIR2004} and \citet{TZ2007} have also used 1-D radiative
transfer calculations (with slightly different numerical methods) to
explore high-redshift ionization structures, with the addition of
self-consistent temperature calculations. \citet{TZ2007} also included
collisional ionization, and an evolving mean IGM density, while
\citet{SIR2004} (who were focused on the evaporation of minihalos by
the I-front) included gas dynamics and trace metals. Like the present
work, \citet{TZ2007} included secondary ionizations, and both papers
included helium, and calculations of the time evolution of the
front. Neither paper included the ionization of hydrogen by helium
recombination radiation, and they explored only blackbody (stellar)
and power-law or truncated power-law (quasar or ``miniquasar'')
sources, rather that the ultra luminous obscured quasars that we are
interested in here (though the miniquasar spectra are fairly similar
to some of our spectra). \citet{SIR2004} found, as we did here, that
I--fronts around sources with harder spectra are thicker.  However,
they studied sources at higher redshift ($z=9$), which were much less
luminous, and the hardest spectrum they examined (a $10^5$ K
blackbody, representing pop III stars) was still softer than most of
the spectra in the range included in the present paper. As a result,
they found fronts much thinner (up to $\sim 0.01$ Mpc; see their
Figures 7 and 8) than most of those discussed in the present paper.
Among the many interesting findings of \citet{TZ2007} are several
relevant to the present study. Even with their harder power-law
spectrum ($s=1$ extending from $200$--$10^4$ eV), they did not find
extended ionization tails outside of the front, due to the combination
of shorter lifetimes and higher redshift (when the IGM density was
higher). They do, however, find an extended kinetic temperature
structure coupled to the spin temperature so as to be observable with
21 cm instruments, and find that such observations could be useful in
discriminating between various source spectra.

Finally, we found that the contours of degeneracy in our parameter
space for measurements using 21 cm observations are oriented
differently from those for measurements using Lyman-series absorption
spectra. This suggests that studies combining both types of observations
have the potential to break the degeneracy and constrain both the
intrinsic absorption column and spectral index of a quasar's ionizing
radiation at $z>6$.

\section{Acknowledgments}

We would like to thank Andrei Mesinger, Stuart Wyithe, and Niel Brandt
for valuable discussions about this work, and Saleem Zaroubi, Martin
Haehnelt, and James Bolton for useful comments on the manuscript. ZH
acknowledges support for this work by NASA through grant NNG04GI88G,
and by the Pol\'anyi Program of the Hungarian Office of Technology. We
thank the referee, Rajat Mani Thomas, for a careful and prompt reading
of the manuscript.

\bibliography{paper1.bib}

\end{document}